 \documentclass[smallextended]{svjour3}       
 \smartqed  
 \usepackage{graphicx}

\usepackage{amsmath,latexsym,amssymb,amsbsy,amsfonts,longtable,lscape,ctable, enumerate, threeparttable,subfigure, mathtools}
\usepackage{algorithm, algorithmic}
\usepackage{natbib}
\usepackage[titletoc,toc,title]{appendix}
\newtheorem{assumption}{Assumption}
\DeclareMathOperator*{\argmin}{argmin} 
\DeclarePairedDelimiter{\norm}{\lVert}{\rVert}

\usepackage[final]{hyperref}
\hypersetup{
  colorlinks,%
  citecolor=black,%
  filecolor=black,%
  linkcolor=black,%
  urlcolor=black
}

\begin{document}
\title{MM for Penalized Estimation}
\author{Zhu Wang}

\institute{Zhu Wang \at
  Department of Population Health Sciences\\
  UT Health San Antonio,
  San Antonio, TX, USA \\
               \email{zhuwang@gmail.com}           
 }

\maketitle

\begin{abstract}
  Penalized estimation can conduct variable selection and parameter estimation simultaneously. The general framework is to minimize a loss function subject to a penalty designed to generate sparse variable selection. The majorization-minimization (MM) algorithm is a computational scheme for stability and simplicity, and the MM algorithm has been widely applied in penalized estimation. Much of the previous work has focused on convex loss functions such as generalized linear models. When data are
    contaminated with outliers, robust loss functions can generate more reliable estimates. Recent literature has witnessed a growing impact of nonconvex loss-based methods, which can generate robust estimation for data contaminated with outliers. This article investigates MM algorithm for penalized estimation, provides innovative optimality conditions and establishes convergence theory with both convex and nonconvex loss functions. With respect to applications, we focus on several nonconvex loss functions, which were formerly studied in machine learning for regression and classification problems. Performance of the proposed algorithms are evaluated on simulated and real data including cancer clinical status. Efficient implementations of the algorithms are available in the \texttt{R} package \texttt{mpath} in CRAN. 

\end{abstract}
\section{Introduction}
In predictive modeling, the data may contain a large number of predictors such as clinical risk factors of hospital patients, or gene expression profiles of cancer patients. It is a common interest to identify effective predictors and estimate predictor parameters simultaneously. A popular approach is to optimize an objective function, log-likelihood for instance, subject to a penalty. Penalty functions including the least absolute shrinkage and
selection operator (LASSO), smoothly clipped absolute deviation (SCAD) and minimax concave penalty (MCP) have been widely applied to many statistical problems \citep{Tibs:1996, fan2001variable, zhang2010nearly}. 

To simplify computations, the majorization-minimization (MM) algorithm provides a useful framework to decomposing a complex minimization problem into a sequence of relatively elementary problems. The MM algorithm has been an interesting research topic in statistics, mathematics and engineering \citep{lange2016mm, byrne2018auxiliary, razaviyayn2013unified}. In particular, penalized regression with MM has attracted some attention, see \citet{schifano2010majorization} and references therein.
The article emphasizes the quadratic majorization (QM) for the generalized linear models.
Despite convergence analysis of the QM algorithm, the loss function was convex. \citet{chi2014robust} developed a QM algorithm to a LASSO type minimum distance estimator for classification, where the limiting behavior relies on more strict assumptions on the penalties: it is an elastic net type of penalty, not the original LASSO, nor the nonconvex penalty SCAD.

In this article, we provide new results for the MM algorithm. The major contributions are three-fold: First, we construct optimality conditions for penalized estimation and penalized surrogate estimation in the MM algorithm. The connection between the two optimization problems are established such that penalty tuning parameters can be conveniently determined from the latter. Second, we establish contemporary theory that the QM algorithm attains the minimum value with convex loss and penalty.
Furthermore, we develop the limiting behavior of the MM algorithm for nonconvex loss functions with the LASSO, SCAD or MCP penalty. Finally, several penalized nonconvex loss functions, for the first time, are applied with QM algorithm for robust regression and classification. 

Nonconvex loss functions play a critical role in robust estimation. Compared to traditional convex loss functions, a nonconvex function may provide more resistant estimation when data are contaminated by outliers. For instance, the least squares loss is not robust to outliers \citep{alfons2013sparse}. Furthermore, variable selection is also strongly influenced by outliers. 
Robust estimation can be obtained by downweighting the impact of the extreme values. Examples include the Huberized LASSO for regression and classification \citep{rosset2007piecewise} and penalized least absolute deviations (LAD) \citep{wang2007robust}. 
However, a breakdown point analysis shows that convex loss functions such as the LAD-LASSO or penalized M-estimators are not robust enough and even a single outlier can send the regression coefficients to infinity \citep{alfons2013sparse}.
On the other hand, bounded nonconvex loss functions have shown more robust to outliers than traditional convex loss functions in regression and classification \citep{wu2007robust, chi2014robust, li2018boosting, wang2018quadratic, wang2018robust}. 
Therefore, this article focuses on applications of penalized nonconvex loss functions. 

The rest of this article is organized as follows. In Section 2, we study optimality conditions of penalized loss functions. In Section 3, we describe the MM algorithm, provide optimality conditions of penalized surrogate loss functions in the MM algorithm, and establish connections of minimizers between the original and surrogate optimization problems. We provide details with applications to
nonconvex loss functions. In Section 4, convergence analysis is conducted. In Section 5, a simulation study
in regression and classification is presented. In Section 6, the proposed methods are evaluated on data application predicting breast cancer clinical status. We conclude with a discussion in Section 7. Some numerical results, technical details such as Clarke subdifferential $\partial_C$ and regular at a point, and proofs are presented in the online supplement.

\section{Penalized estimation}
For training data $(x_{ij}, y_i), i=1, ..., n, j=0, 1, ..., p$, let $x_i=(x_{i0}, ..., x_{ip})^\intercal$ denote a $(p+1)$-dimensional predictor with the first entry 1, $(y_1, ..., y_n)^\intercal$ is the response vector, $\phi=(\beta_0, \beta_1, ..., \beta_p)^\intercal$ is a $(p+1)$-dimensional coefficient vector with $\beta_0$ the intercept. 
Consider a loss function $\Gamma(u)$, for instance, $\Gamma(u)=\norm{u}^2, u=y-f$ in linear regression, where $f$ is the prediction of $y$. 
For some link function $\psi(\cdot)$ define
\begin{equation}\label{eqn:mloss2}
	L(\phi)=\frac{1}{n}\sum_{i=1}^nL_i(\phi), L_i(\phi)=\Gamma(\psi(x_i^\intercal\phi)).
\end{equation}
The goal is to minimize the empirical loss $L(\phi)$.
For the purpose of variable selection, define a penalized loss function $F: \mathbb{R}^{p+1} \to \mathbb{R}$
\begin{equation}\label{eqn:plik}
	F(\phi)\triangleq
	L(\phi)+ 
  \sum_{j=1}^{p} \left(\alpha p_\lambda(|\beta_j|) + \lambda\frac{1-\alpha}{2}\beta_j^2\right) , 
\end{equation}
where $\lambda > 0, 0 \leq \alpha \leq 1$, and $p_\lambda(|\beta_j|)$ is the penalty function 
satisfying the following assumptions \citep{schifano2010majorization}:
\begin{assumption}\label{ass:pen}
    $p_\lambda(\theta)$ is continuously differentiable, nondecreasing and concave on $(0, \infty)$ with $p_\lambda(0)=0$ and $0 < p'_\lambda(0+) < \infty$.
\end{assumption}

Minimizing the penalized loss function $F(\phi)$ can provide shrinkage estimates thus leading to zero values for small coefficients in magnitude. 
    The following penalty functions have been widely studied:

(i) the LASSO penalty \citep{Tibs:1996}, for $\lambda \geq 0$
\begin{equation*}\label{eqn:enet}
	p_\lambda(|\theta|)=\lambda|\theta|.
\end{equation*}

(ii) the SCAD penalty \citep{fan2001variable}
\begin{equation}\label{eqn:scad}
  \begin{aligned}
	  p'_\lambda(\theta)&=\lambda\left\{I( \theta < \lambda)+\frac{(a\lambda-\theta)_+}{(a-1)\lambda} I(\theta \geq \lambda)\right\}
  \end{aligned}
\end{equation}
for $\lambda \geq 0$ and $a > 2$, where $p'(\cdot)$ is a derivative,
and $t_+$ is the positive part of $t$. 

(iii) the MCP penalty \citep{zhang2010nearly}, for $\lambda \geq 0$ and $a > 1$
\begin{equation}\label{eqn:mcp}
  \begin{aligned}
	  p'_\lambda(\theta)&=
    (\lambda - \theta/a)I(\theta \leq a\lambda).
  \end{aligned}
\end{equation}

The SCAD and MCP penalty functions with large value $a$ are comparable to the LASSO since $p'_\lambda(\theta) \to \lambda$ if $a \to \infty$. These penalty functions satisfy Assumption~\ref{ass:pen} and Assumption~\ref{ass:pen2} (more restrictive than Assumption~\ref{ass:pen}).
\begin{assumption}\label{ass:pen2}
    $p_\lambda(\theta)$ is continuously differentiable, nondecreasing and concave on $(0, \infty)$ with $p_\lambda(0)=0$, $0 < p'_\lambda(0+) < \infty$ and directional derivative at $0$ in the direction $\varepsilon$ given by $p'_\lambda(0; \varepsilon)=\zeta\lambda|\varepsilon|$ for some $\zeta > 0$. 
\end{assumption}
In particular, $\zeta=1$ for the LASSO, SCAD and MCP penalty. Assumption~\ref{ass:pen2} also covers hard thresholding penalty.

Assumption~\ref{ass:pen} ensures the penalty function is locally Lipschitz continuous at every point including the origin due to a bounded derivative.
A penalty function may violate Assumption~\ref{ass:pen}. For instance, the bridge penalty $p(\theta)=\theta^{a}$ for $a \in (0, 1)$ has $p'_\lambda(0+) = \infty$. In other words, the bridge penalty is not Lipschitz continuous near $\theta=0$. 
    
    We establish conditions for minimizing $F(\phi)$. Let $\phi^\ast=(\beta^\ast_0, \beta^\ast_1..., \beta^\ast_p)^\intercal$ denote a minimizer and corresponding value $\beta^\ast=(\beta^\ast_1, ..., \beta^\ast_p)$ without the intercept. Since $\beta_0$ is not penalized, we focus on optimality conditions on $\beta^\ast$, since an optimality condition on $\beta_0$ is a trivial problem in the current setting. Whether $\phi^\ast$ is a local or global minimizer is specified in the context. If
    $\phi^\ast$ is a minimizer, then usually a set of conditions constraining $\phi^\ast$ must be satisfied. If these conditions are comprehensively constructed, we call them complete optimality conditions, which is the typical case when dealing with convex optimization problems. On the other hand, if only a subset of complete optimality conditions are known, we call them incomplete optimality conditions. This would occur in nonconvex optimization problems. 
\begin{theorem}\label{thm:opt}
    Suppose $\phi^\ast$ is a local minimizer of $F(\phi)$, where $L(\phi)$ is continuously differentiable.
    Then:
    \begin{enumerate}[(a)]
        \item Under Assumption~\ref{ass:pen}, optimality conditions for $\beta^\ast$ 
    are given by
    \begin{equation}\label{eqn:opt1}
        \frac{1}{n}\sum_{i=1}^n\frac{\partial{L_i(\phi^\ast)}}{\partial\beta_j}+\alpha \text{sign}(\beta_j^\ast)p'_{\lambda}(\beta^\ast_j)+\lambda(1-\alpha)\beta^\ast_j=0, \text{ if } \beta^\ast_j \neq 0,\\    
        \end{equation}
    \begin{equation}\label{eqn:opt1.2}
            \frac{1}{n}\sum_{i=1}^n\frac{\partial{L_i(\phi^\ast)}}{\partial\beta_j}+\alpha
            \tau=0, \tau \in \partial_Cp_\lambda(0), \text{ if } \beta^\ast_j=0,
        \end{equation}
            for $j=1, ..., p$.
\item  Under Assumption~\ref{ass:pen2}, an optimality condition for $\beta^\ast_j = 0, j=1, ..., p$ is given by
    \begin{equation}\label{eqn:opt2}
        \frac{1}{n}\sum_{i=1}^n\frac{\partial{L_i(\phi^\ast)}}{\partial\beta_j}+\alpha\zeta\lambda e=0, e \in [-1, 1]. 
    \end{equation}
            If $p_\lambda(|\theta|)$ is also regular at $0$, the condition is complete. 
    \end{enumerate}
\end{theorem}

For the LASSO, SCAD and MCP penalty functions with $\zeta=1$, from (\ref{eqn:opt2}) we define
    \begin{equation}\label{eqn:opt3}
        \lambda_{\max}\triangleq \max_{1\leq j\leq p}\frac{1}{n\alpha}\left|\sum_{i=1}^n\frac{\partial{L_i(\phi^\ast)}}{\partial\beta_j}\right|.
    \end{equation}
In real data analysis, we often need to determine a good value $\lambda$, which can be chosen from select values closely related to the quantity $\lambda_{\max}$. 
From Theorem~\ref{thm:opt}, if $\beta_j=0, j=1, ..., p$, we deduce that $\lambda \geq \lambda_{\max}$. 
On the other hand, the converse is not valid, unless stronger conditions are posed as in the next theorem.

\begin{theorem}\label{thm:opt1.1}
    Assume that $L(\phi$) is differentiable convex, and $p_\lambda(|\theta|)$ is the LASSO penalty function. Then $\phi^\ast$ is a global minimizer of $F(\phi)$ if and only if
\begin{equation*}\label{eqn:clarke01.01}
    0 \in \partial F(\phi^\ast)=\left\{\nabla L(\phi^\ast)\right\}+\partial\sum_{j=1}^p\alpha p_\lambda(|\beta_j^\ast|)+\left\{\lambda(1-\alpha)(0, \beta^\ast)^\intercal\right\},
\end{equation*}
    where $\partial F(\phi^\ast)$ is the subdifferential of $F(\phi)$ at $\phi=\phi^\ast$.
    Or equivalently
    \begin{equation}\label{eqn:opt2.1}
    \frac{1}{n}\sum_{i=1}^n\frac{\partial{L_i(\phi^\ast)}}{\partial\beta_j}+\alpha
    \text{sign}(\beta_j^\ast)p'_{\lambda}(\beta^\ast_j)+\lambda(1-\alpha)\beta^\ast_j=0, \text{ if } \beta^\ast_j \neq 0,
    \end{equation}
    \begin{equation}\label{eqn:opt2.1_1}
        \frac{1}{n}\sum_{i=1}^n\frac{\partial{L_i(\phi^\ast)}}{\partial\beta_j}+\alpha\lambda e=0, e \in [-1, 1], \text{ if } \beta^\ast_j=0,
        \end{equation}
            for $j=1, ..., p$.
    \end{theorem}

The optimality condition in Theorem~\ref{thm:opt1.1} is standard for convex functions, yet helpful when investigating the relationship between the function $F(\phi)$ and its surrogate function in the MM algorithm described below. The value $\lambda_{\max}$ has much stronger implications under Theorem~\ref{thm:opt1.1} due to the convexity assumption. From(\ref{eqn:opt2.1_1}) Theorem~\ref{thm:opt1.1} concludes that $\lambda_{\max}$ is the smallest $\lambda$ such that $\beta_j=0, j=1, ..., p$.
Without convexity assumption, since (\ref{eqn:opt2}) is not sufficient and/or incomplete, the value $\lambda_{\max}$ may neither guarantee $\beta_j=0$ nor produce the smallest $\lambda$ with $\beta_j=0$.

\section{MM algorithms}\label{sec:mm}
\subsection{Penalized estimation using MM algorithms}\label{sec:mmpen}
We say function $\gamma(u|z)$ majorizes function $\Gamma(u)$ at $z$ if the following holds $\forall\ u, z \in \mathbb{R}^m$: 
\begin{equation}\label{eqn:mm0}
	\Gamma(u) \leq \gamma(u|z), \quad \Gamma(z)=\gamma(z|z).
\end{equation}
The MM algorithm is an iterative procedure. Given an estimate $z=u^{(k)}$ in the $k$th iteration, $\gamma{(u|z)}$ is minimized at the $k+1$ iteration to obtain an updated minimizer $u^{(k+1)}$.
The MM algorithm then updates $z$ with the new estimate $u^{(k+1)}$. This process is repeated until convergence. 
From (\ref{eqn:mm0}) the MM algorithm generates a descent sequence of estimates:
\begin{equation*}\label{eqn:mm}
  \Gamma(u^{(k+1)}) \leq 
	\gamma(u^{(k+1)}|u^{(k)}) \leq 
	\gamma(u^{(k)}|u^{(k)})  =  
  \Gamma(u^{(k)}). 
\end{equation*}
This relationship leads to the following result.
\begin{lemma}\label{lem:mm}
    Suppose $\gamma(u|z)$ majorizes $\Gamma(u)$ at $z$. If $u^\ast$ is a local (global) minimizer of $\Gamma(u)$, $u^\ast$ is a local (global) minimizer of $\gamma(u|u^\ast)$. 
\end{lemma}

The converse of Lemma~\ref{lem:mm} is not true. Consider $\gamma(u|u^\ast)=u^2, u^\ast=0$. Let $\Gamma(u)= -u^2$. Therefore $\gamma(u|z)$ majorizes $\Gamma(u)$, 
$u^\ast=\argmin \gamma(u|u^\ast)$ and $\nabla \Gamma(u^\ast)=\nabla \gamma(u^\ast|u^\ast)=0$. However, the converse is invalid. To overcome this problem, for differentiable convex functions, we develop a connection between a function and its ``surrogate'' function in a broader sense, including but not limited to the induced functions by majorization. 
\begin{lemma}\label{lem:mm2}
    Suppose $\Gamma(u)$ and $\gamma(u|z)$ are differentiable convex functions. Then $\phi^\ast$ is a global minimizer of $\Gamma(u)$ satisfying $\nabla\Gamma(u^\ast)=\nabla\gamma(u^\ast|u^\ast)$ if and only if $\phi^\ast$ is a global minimizer of $\gamma(u|u^\ast)$.
\end{lemma}
To accommodate predictors, define 
\begin{equation}\label{eqn:ell}	
    \ell(\phi|\phi^{(0)})=\frac{1}{n}\sum_{i=1}^n\ell_i(\phi|\phi^{(0)}), \ \ell_i(\phi|\phi^{(0)})=\gamma(\psi(x_i^\intercal\phi)|\phi^{(0)}).
\end{equation}
Obviously the transformation $\psi(\cdot)$ preserves the relationship in (\ref{eqn:mm0}) such that $\ell_i(\phi|\phi^{(0)})$ majorizes $L_i(\phi)$ at the current estimate $\phi^{(0)}=(\beta_0^{(0)}, \beta_1^{(0)}, ..., \beta_p^{(0)})^\intercal$. 
Along the line of Lemma~\ref{lem:mm2}, we characterize a class of surrogate functions: 
\begin{assumption}\label{ass:sur}
    Two differentiable functions have the same derivative at point $\phi$: 
    \begin{equation*}\label{eqn:eqder}
        \nabla L(\phi)=\nabla\ell(\phi|\phi).
    \end{equation*}
    \end{assumption}
Consider the following surrogate penalized loss:
\begin{equation}\label{eqn:plik1}
	Q(\phi|\phi^{(0)})\triangleq
	\ell(\phi|\phi^{(0)})+ 
	\sum_{j=1}^{p} \left(\alpha p_\lambda(|\beta_j|) + \lambda\frac{1-\alpha}{2}\beta_j^2\right).
\end{equation}
From Lemma~\ref{lem:mm} we deduce that if $\phi^\ast$ is a minimizer of $F(\phi)$, it must be a minimizer of $Q(\phi|\phi^\ast)$. The following results lay out not only the optimality conditions of the surrogate function, but also the connection between the surrogate and its original function deduced from. 
\begin{theorem}\label{thm:clarkeequ}
    Assume that $L(\phi)$ and $\ell(\phi|\phi^{(0)})$ are continuously differentiable functions, and $\ell(\phi|\phi^{(0)})$ majorizes $L(\phi)$ at $\phi^{(0)}$. 
    Suppose $\phi^\ast$ is a local minimizer of $F(\phi)$, and Assumption~\ref{ass:sur} holds at $\phi^\ast$. Then:
    \begin{enumerate}[(a)]
        \item        Under Assumption~\ref{ass:pen}, an optimality condition is 
    \begin{equation*}
        0 \in \partial_C Q(\phi^\ast|\phi^\ast) = \partial_C F(\phi^\ast). 
    \end{equation*}
\item  Under Assumption~\ref{ass:pen2}, an optimality condition for $\beta^\ast_j = 0, j=1, ..., p$ is
    \begin{equation}\label{eqn:thm3.1}
        \frac{1}{n}\sum_{i=1}^n\frac{\partial{\ell_i(\phi^\ast|\phi^\ast)}}{\partial\beta_j}+\alpha\zeta\lambda e=0, e \in [-1, 1]. 
    \end{equation}
If $p_\lambda(|\theta|)$ is also regular at $0$, the condition is complete. 
    \end{enumerate}
\end{theorem}

For the LASSO, SCAD and MCP penalty functions, from (\ref{eqn:thm3.1}) we define
    \begin{equation}\label{eqn:thm3.4}
        \tilde\lambda_{\max}\triangleq \max_{1\leq j \leq p}\frac{1}{n\alpha}\left|\sum_{i=1}^n\frac{\partial{\ell_i(\phi^\ast|\phi^\ast)}}{\partial\beta_j}\right|.
    \end{equation}
Notice $\tilde\lambda_{\max}$ is the same as $\lambda_{\max}$ in (\ref{eqn:opt3}) under the specified assumptions. Theorem~\ref{thm:clarkeequ} implies that if $\beta_j=0, j=1, ..., p$, then $\lambda \geq \tilde\lambda_{\max}$. Like before, the converse is generally invalid unless stronger conditions are provided as in the next theorem.
\begin{theorem}\label{thm:mmopt2}
    Assume that $L(\phi)$ and $\ell(\phi|\phi^{(0)})$ are differentiable convex functions, 
    and $p_\lambda(|\theta|)$ is the LASSO penalty function, 
    Then $\phi^\ast$ is a global minimizer of $F(\phi)$, where Assumption~\ref{ass:sur} holds at $\phi^\ast$, 
    if and only if $\phi^\ast$ is a global minimizer of $Q(\phi|\phi^\ast)$ or equivalently 
    \begin{equation}\label{eqn:opt4.1}
    \frac{1}{n}\sum_{i=1}^n\frac{\partial{\ell_i(\phi^\ast|\phi^\ast)}}{\partial\beta_j}+\alpha
    \text{sign}(\beta_j^\ast)p'_{\lambda}(\beta^\ast_j)+\lambda(1-\alpha)\beta^\ast_j=0, \text{ if } \beta^\ast_j \neq 0,
    \end{equation}
    \begin{equation}\label{eqn:opt4.2}
        \frac{1}{n}\sum_{i=1}^n\frac{\partial{\ell_i(\phi^\ast|\phi^\ast)}}{\partial\beta_j}+\alpha\lambda e=0, e \in [-1, 1], \text{ if } \beta^\ast_j = 0. 
    \end{equation}
\end{theorem}

Theorem~\ref{thm:mmopt2} on the penalized surrogate loss $Q(\phi|\phi^{(0)})$ is a mirror of Theorem~\ref{thm:opt1.1} on the penalized loss $F(\phi)$. Furthermore, it is reasonable to deduce that similar results hold for the EM algorithm (a special case of the MM algorithm) in penalized estimation. 
Such a connection, due to the convexity assumption, is stronger than that between Theorem~\ref{thm:opt} and Theorem~\ref{thm:clarkeequ}. 
These results suggest that instead of minimizing the original objective function $F(\phi)$, we can minimize the surrogate function $Q(\phi|\phi^{(0)})$. The iterative MM procedure is summarized in Algorithm~\ref{alg:alg1}.
\begin{algorithm}
	\begin{algorithmic}[1]
	\caption{MM Algorithm}\label{alg:alg1}
	\STATE \textbf{Initialize} $\phi^{(0)}$ and set $k=0$
	    \REPEAT
	    \STATE Compute $\phi^{(k+1)}= \argmin Q(\phi|\phi^{(k)})$
		    \STATE $k= k+1$
			    \UNTIL convergence of $\phi^{(k)}$
\end{algorithmic}
\end{algorithm}
\subsection{Quadratic majorization}\label{sec:qm}
The quadratic majorization (QM) 
is based on the second-order Taylor expansion for a twice differentiable function $\Gamma(u)$ such that
\begin{equation*}\label{eqn:ta}
  \begin{aligned}
	  \Gamma(u)&=\Gamma(z) + (u-z)^\intercal\nabla \Gamma(z) + \frac{1}{2}(u-z)^\intercal \nabla^2\Gamma(w)(u-z),\\ 
  \end{aligned}
\end{equation*}
for some $w_i$ on the line segment connecting $z$ and $u$.
Suppose $\tilde{B}$ is positive definite and $\tilde{B}-\nabla^2 \Gamma(w)$ is semipositive definite for all arguments $w$, then 
\begin{equation}\label{eqn:ma}
  \begin{aligned}
      \gamma(u|z)\triangleq \Gamma(z) + (u-z)^\intercal \nabla \Gamma(z) + \frac{1}{2}(u-z)^\intercal \tilde{B} (u-z)\\
  \end{aligned}
\end{equation}
majorizes $\Gamma(u)$ at $z$. 
Since $\nabla\gamma(z|z)=\nabla\Gamma(z)$, Assumption~\ref{ass:sur} holds at $z$ for functions $\gamma(u|z)$ and $\Gamma(u)$.
The QM is a special application of the more general MM algorithm when a suitable bound $\tilde{B}$ exists \citep{lange2016mm}. The generic terms defined by $\ell(\phi|\phi^{(0)})$ in (\ref{eqn:ell}) can be replaced with the express formula (\ref{eqn:ma}). 

While an upper bound $\tilde{B}$ may not be unique, to simplify computation as often possible, assume $\tilde{B}$ is a square diagonal matrix with all its main diagonal entries equal, i.e., $\tilde{B}=BI$, where $B$ is a scalar and $I$ is the identity matrix. 
Furthermore the link function $\psi(\cdot)$ takes different forms:
\begin{equation}\label{eqn:psi}
	\psi(f)=
	\begin{cases}
            y-f, &\text{ for regression,}\\
            yf,  &\text{ for classification}.  
  \end{cases}
\end{equation}
Define $u=y-f$ in regression problems. By (\ref{eqn:ma}) and the chain rule, we have
\begin{equation*}
	\gamma(u|z)=\Gamma(y-z) - \nabla \Gamma(y-z)(f-z) + \frac{B}{2}(f-z)^\intercal(f-z).
\end{equation*}
For classification problems with margin $u=yf, y\in \{-1, 1\}$, we have
\begin{equation*}
	\gamma(u|z)=\Gamma(yz)+y\nabla \Gamma(yz)(f-z)+\frac{B}{2}(f-z)^\intercal (f-z).
\end{equation*}
With $f_i=x_i^\intercal \phi, z_i=x_i^\intercal \phi^{(0)}$, minimizing unpenalized loss $L(\phi)$ in (\ref{eqn:mloss2}) amounts to minimizing 
\begin{equation}\label{eqn:l2}
    \ell(\phi|\phi^{(0)})=\frac{B}{2n}\sum_{i=1}^n(h_i(\phi^{(0)})-x_i^\intercal \phi)^2,
\end{equation}
where
\begin{equation}\label{eqn:hval}
      h_i(\phi^{(0)})=
  \begin{cases} 
      \frac{\nabla L_i(y_i-z_i)}{B}+z_i, &\mbox{for regression}, \\
      -y_i \frac{\nabla L_i(y_i z_i)}{B} + z_i, &\mbox{for classification.}
  \end{cases}
\end{equation}
The constant weight $B$ has no impact on the estimation given $h_i(\phi^{(0)})$, in contrast to the penalized estimation described below. Putting together, with iterated least squares estimation for the updated response variable, Algorithm~\ref{alg:alg2} can be used to minimize $L(\phi)$ in (\ref{eqn:mloss2}) and estimate parameters.
\begin{algorithm}
	\begin{algorithmic}[1]
	\caption{QM Algorithm}\label{alg:alg2}
	\STATE \textbf{Initialize} $\phi$
	    \REPEAT
        \STATE Compute $z_i=x_i^\intercal\phi$ and $h_i(\phi)$ in (\ref{eqn:hval})
		\STATE Minimize least squares (\ref{eqn:l2}) to update $\phi$
			    \UNTIL convergence of $\phi$
\end{algorithmic}
\end{algorithm}

    For penalized estimation, the surrogate loss function (\ref{eqn:plik1}) simplifies to:
\begin{equation}\label{eqn:plik2} 
	Q(\phi|\phi^{(0)})=
    \frac{B}{2n}\sum_{i=1}^n(h_i(\phi^{(0)})-x_i^\intercal \phi)^2 +
  \sum_{j=1}^{p} \left(\alpha p_\lambda(|\beta_j|) + \lambda\frac{1-\alpha}{2}\beta_j^2\right). 
\end{equation}
 Since the surrogate function $\ell(\phi|\phi^{(0)})$ satisfies Assumption~\ref{ass:sur}, Theorem~\ref{thm:clarkeequ} is applicable. Theorem~\ref{thm:mmopt2} is effective for convex loss functions such as those in generalized linear models (GLM), where quadratic majorization may result in possibly different form than (\ref{eqn:hval}). 
 Minimizing $F(\phi)$ in (\ref{eqn:plik}) can be achieved with Algorithm~\ref{alg:alg3}, and technical details are discussed in the subsequent context and supplement. 
\begin{algorithm}
	\begin{algorithmic}[1]
	\caption{QM Algorithm for Penalized Estimation}\label{alg:alg3}
	\STATE \textbf{Initialize} $\phi$
	    \REPEAT
        \STATE Compute $z_i=x_i^\intercal\phi$ and $h_i(\phi)$ in (\ref{eqn:hval}). 
	    \STATE 
	Minimize penalized least squares (\ref{eqn:plik2}) 
		to update $\phi$
			    \UNTIL convergence of $\phi$
\end{algorithmic}
\end{algorithm}

\subsection{Penalized least squares}\label{sec:pl2}
Minimizing the penalized least squares (\ref{eqn:plik2}) is hampered by the fact that the penalty function $p_\lambda(|\cdot|)$ is not differentiable at $0$. Let $G(\cdot|\cdot)$ be a function that approximates the penalty function $p_\lambda(|\cdot|)$, the following methods have been proposed in the literature.
\begin{enumerate}[(i)]
	\item
        No approximation is needed \citep{Frie:regu:2010, breheny2011coordinate}:
	\begin{equation}\label{eqn:nomm}
		G(\theta|\theta^{(0)})\triangleq p_\lambda(|\theta|).
\end{equation}
	\item
		\citet{zou2008one} studied a local linear approximation (LLA): 
\begin{equation}\label{eqn:lla}
	G(\theta|\theta^{(0)})\triangleq p_\lambda(|\theta^{(0)}|)+(|\theta|-|\theta^{(0)}|)p'_\lambda(|\theta^{(0)}|+).
\end{equation}
	\item
\citet{fan2001variable, hunter2005variable} considered a local quadratic approximation (LQA): 
\begin{equation}\label{eqn:lqa}
	G(\theta|\theta^{(0)})\triangleq p_\lambda(|\theta^{(0)}|)+\frac{(\theta^2-\theta^{(0)2})p'_\lambda(|\theta^{(0)}|+)}{2|\theta^{(0)}|},
\end{equation}
where $p'_\lambda(|\theta^{(0)}|+)=\lim\limits_{\tilde\theta \to |\theta^{(0)}|+}p_\lambda'(\tilde\theta)$.
\end{enumerate}

In these proposals $G(\theta|\theta^{(0)})$ essentially majorizes $p_\lambda(|\theta|)$ at $\theta^{(0)}$. We implement method (i) in the algorithm since the efficient coordinate decent algorithm can easily provide penalized least squares estimates for a variety of penalty functions including the LASSO, SCAD and MCP \citep{Frie:regu:2010, breheny2011coordinate}. We nevertheless investigate properties of all three approaches in Section~\ref{sec:conv}. 

\subsection{Applications to nonconvex loss functions}\label{sec:robust}
\begin{figure}[!htbp]
  \begin{center}
      \includegraphics[scale=0.3]{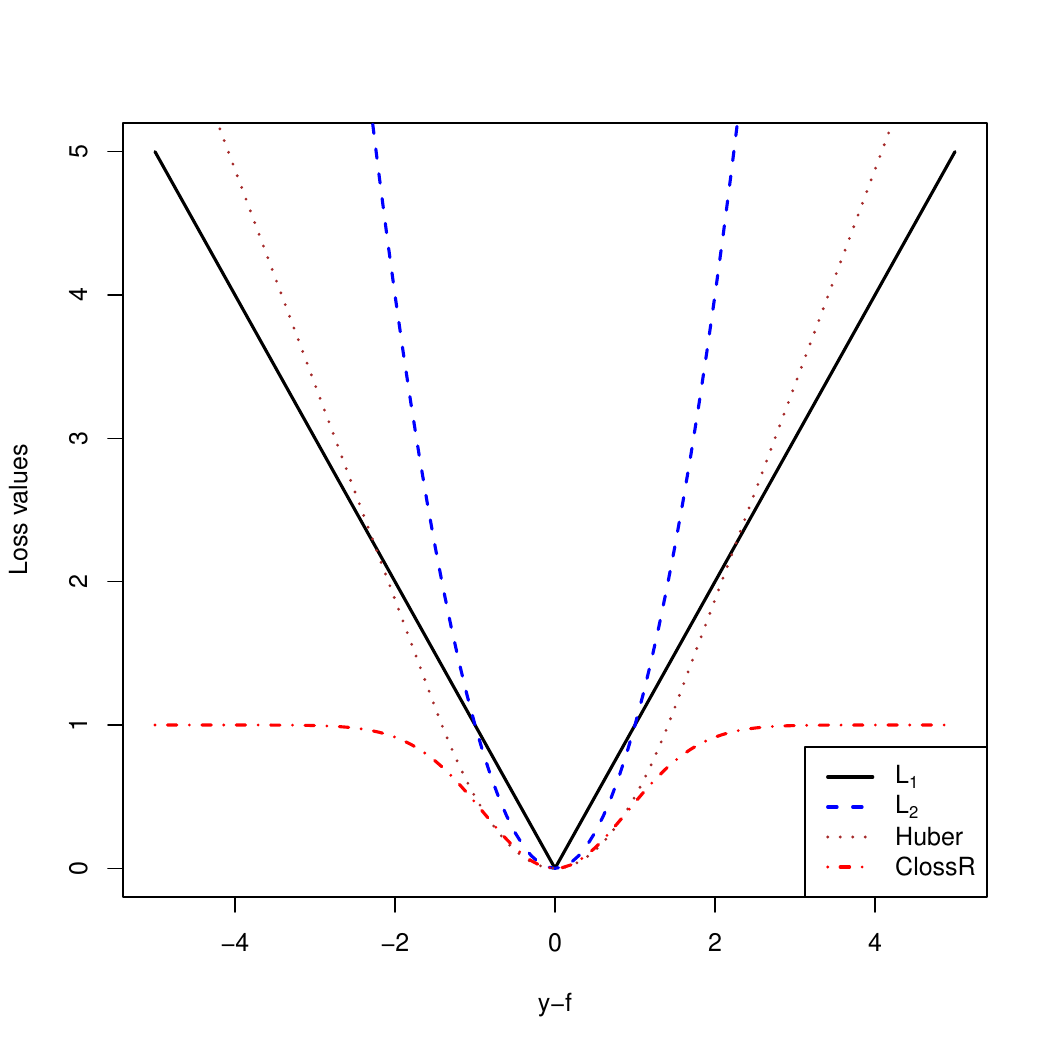}
    \caption{Loss functions for regression}
    \label{fig:loss2}
  \end{center}
\end{figure}

Traditional regression is based on minimizing a loss function $\Gamma(u)$, where $\Gamma(u)=\norm{u}^2, u=y-f$. 
The $L_2$ loss is not robust to outliers. Robust regression has utilized absolute difference $\Gamma(u)=|u|$ ($L_1$ loss) and Huber loss, both are convex loss functions. The nonconvex loss ClossR with a parameter $\sigma$ is described in Table~\ref{tab:tab1} \citep{wang2018quadratic}.
The ClossR with $\sigma=0.9$ is plotted in Figure~\ref{fig:loss2}, along with the aforementioned loss functions in regression. It is clear that bounded ClossR downweights large values more than other convex loss functions, effectively having better resistance to outliers.  

\begin{figure}[!htbp]
  \begin{center}
      \includegraphics[scale=0.3]{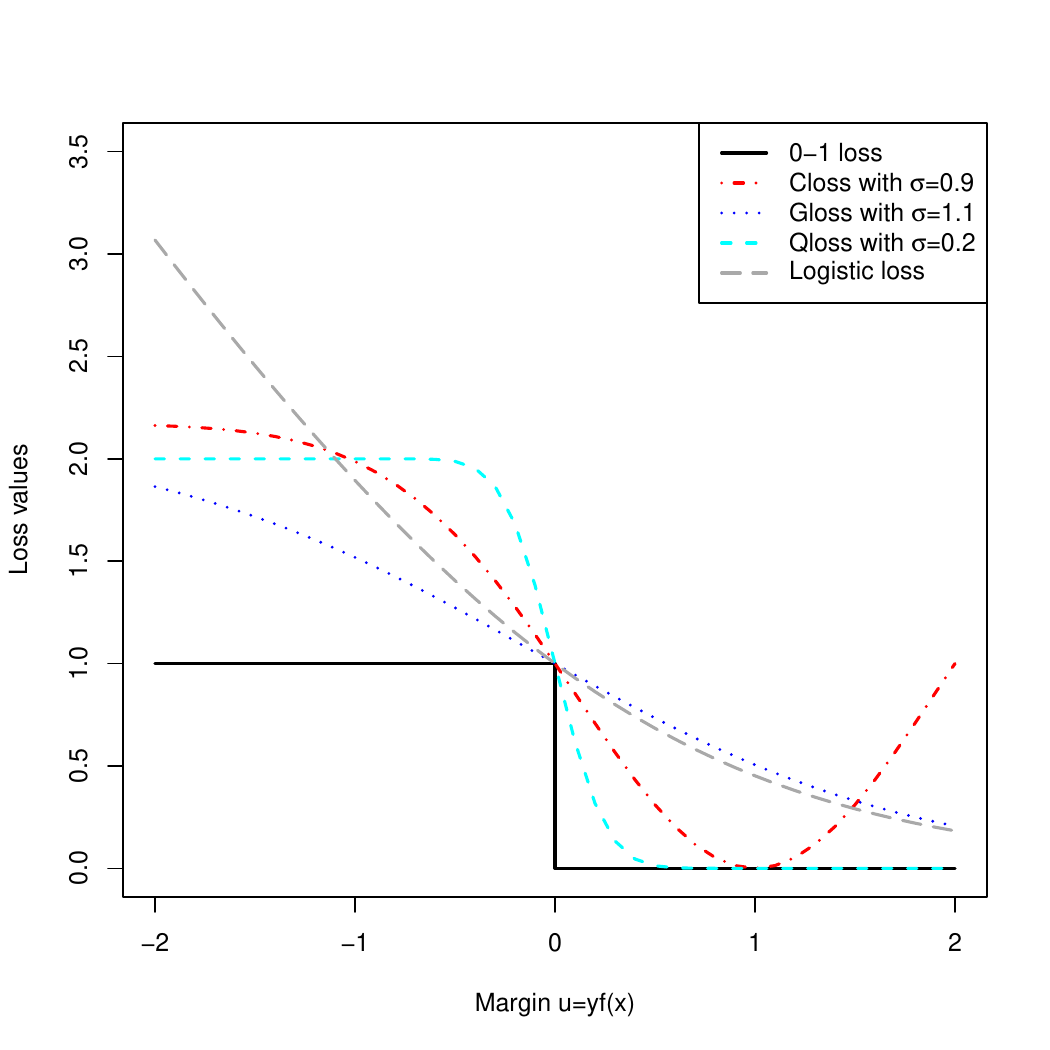}
    \caption{Loss functions for classification}
    \label{fig:loss}
  \end{center}
\end{figure}

For a binary outcome $y$ taking values $+1$ and $-1$, denote by $u=yf$ the margin of a classifier $f$. Used in a logistic regression, $\Gamma(u)=\log_2(1+\exp(-u))$ is a popular choice in statistics. However, the logistic loss is unbounded and can't control outliers well. 
Nonconvex loss functions have been proposed to address the robustness of classification, and three examples are listed in Table~\ref{tab:tab1}. These loss functions approximate the 0-1 loss, as shown in Figure~\ref{fig:loss}. 
The $\sigma$ value controls robustness of a loss function. For Closs, a smaller value is more robust to outliers while for other loss functions, a larger value is   more robust. A typical strategy in practice is working from less robust to more robust and selecting a proper value of $\sigma$       according to prediction.

For the loss functions in Table~\ref{tab:tab1}, we can obtain penalized estimates with Algorithm~\ref{alg:alg3}, where factor $B$ has been derived by \citet{wang2018quadratic}. Since the functions are continuously differentiable, Theorem~\ref{thm:opt} and Theorem~\ref{thm:clarkeequ} are applicable. We list $\lambda_{\max}$ in the supplement Table~\ref{tab:tab3}. 
The starting estimates of $\phi$ are based on a functional gradient decent boosting algorithm \citep{wang2018robust}. 
\section{Convergence analysis}\label{sec:conv}
To conduct a convergence analysis of Algorithm~\ref{alg:alg1} for penalized estimation in convex and nonconvex problems, we construct a unified form of majorizers for the loss functions in Section~\ref{sec:robust} as well as those in the GLM. Consider a twice differentiable function $L_i(\phi)$ in (\ref{eqn:mloss2}). The Taylor expansion at a given value $\phi^{(k)}$ is 
\begin{equation*}
	L_i(\phi)=L_i(\phi^{(k)})+(\phi-\phi^{(k)})^\intercal \nabla L_i(\phi^{(k)})+\frac{1}{2}(\phi-\phi^{(k)})^\intercal\nabla^2 L_i(\tilde \phi)(\phi-\phi^{(k)})
\end{equation*}
for some $\tilde\phi$ on the line segment connecting $\phi$ and $\phi^{(k)}$, where
\begin{equation*}
	\nabla L_i(\phi)= \nabla \Gamma\left(\psi(x_i^\intercal\phi)\right) \nabla\psi(x_i^\intercal\phi) x_i,
\end{equation*}
\begin{equation*}
	\begin{aligned}
		\nabla^2 L_i(\phi)&= \left\{\nabla^2 \Gamma\left(\psi(x_i^\intercal\phi)\right) \left(\nabla\psi(x_i^\intercal\phi)\right)^2 +
	                          \nabla \Gamma\left(\psi(x_i^\intercal\phi)\right)\nabla^2\psi(x_i^\intercal\phi)\right\}
				  x_i x_i^\intercal\\
				  &\triangleq d_i(\phi) x_i x_i^\intercal.
	\end{aligned}
\end{equation*}
For $\psi(f)$ defined in (\ref{eqn:psi}), we have 
\begin{equation*}
	d_i(\phi)=\nabla^2 \Gamma\left(\psi(x_i^\intercal\phi)\right),
\end{equation*}
which can be applied to functions including but not limited to the nonconvex functions in Section~\ref{sec:robust}. 
For the GLM, the negative log-likelihood function is
\begin{equation*}
    \Gamma(\phi)=-\left(y\delta - b(\delta)\right)/a(\kappa) - c(y, \kappa) ,
\end{equation*}
where $a(\cdot), b(\cdot)$ and $c(\cdot)$ are known functions, $\delta$ is the canonical parameter  with $\delta_i = \psi(x_i^\intercal\phi)$ (if the link is canonical, $\delta_i = x_i^\intercal\phi)$, and $\kappa$ is the known dispersion parameter. 
The negative log-likelihood function has the form:
\begin{equation*}
	L(\phi)= \frac{1}{n}\sum_{i=1}^n L_i(\phi)\triangleq\frac{1}{n}\sum_{i=1}^n \left\{[-y_i\psi(x_i^\intercal\phi) + b\left(\psi(x_i^\intercal\phi)\right)] /a(\kappa) - c(y_i, \kappa)\right\}.
\end{equation*}
We have
\begin{equation*}
	\nabla L_i(\phi)=\frac{1}{a(\kappa)}\big(-y_i\nabla\psi(x_i^\intercal \phi)+\nabla b(\psi(x_i^\intercal \phi)) \nabla\psi(x_i^\intercal \phi)\big)x_i
\end{equation*}
\begin{equation*}\label{eqn:glm2}
	\begin{aligned}
		\nabla L_i^2(\phi)&=
		\frac{1}{a(\kappa)}\left(-y_i\nabla^2\psi(x_i^\intercal \phi)+\nabla^2 b(\psi(x_i^\intercal \phi))\left(\nabla\psi(x_i^\intercal\phi)\right)^2 +\nabla b(\psi(x_i^\intercal \phi))\nabla^2\psi(x_i^\intercal \phi)\right)
		x_ix_i^\intercal.\\
							      &\triangleq d_i(\phi) x_i x_i^\intercal.
	\end{aligned}
\end{equation*}
For the canonical link $\psi(z)=z$, we have
\begin{equation*}
	d_i(\phi)=\frac{1}{a(\kappa)}\nabla^2 b\left(x_i^\intercal \phi\right).
\end{equation*}
Let $X$ denote the $n \times (p+1)$ matrix of $x_i$ values, $\Omega$ a $n\times n$ diagonal matrix of weights with $i$th diagonal element $d_i(\phi)$. 
Then we have
\begin{equation*}
    \nabla^2L(\phi)=\frac{1}{n}\sum_{i=1}^n\nabla^2 L_i(\phi)=\frac{1}{n}X^\intercal \Omega X.
\end{equation*}
A matrix $W$ is constructed to majorize $L(\phi)$.
Let $\tilde{d}_j(\phi), j=0, 1, ..., p$ denote the eigenvalues of $\nabla^2L(\phi)$.
Assume that 
\begin{equation}\label{eqn:omega}
\varrho=
	\max_\phi\max_{j=0, 1, ..., p}{\tilde{d}_j(\phi)} < \infty,
\end{equation}
and let $W=I$, where $I$ is the identity matrix. 
If $X$ is full rank, we may define
\begin{equation}\label{eqn:omega1}
\varrho=
    \max_\phi\max_{i=1, ..., n}{d_i(\phi)} < \infty, W=\frac{1}{n}X^\intercal X.
    \end{equation}
Obviously the $(p+1) \times (p+1)$ matrix $W$ is positive definite, and all the eigenvalues of $W$ are positive. 
Let $\omega \geq \varrho, \ \omega > 0.$ We have
\begin{equation}\label{eqn:wmat1}
	\omega W - \nabla^2 L(\phi) \succeq 0 I,
\end{equation}
where  the inequality $\succeq$ means that  $\omega W - \nabla^2 L(\phi) - 0 I$ is positive semidefinite.
Denote
\begin{equation}\label{eqn:thm1.38}
	\ell(\phi|\phi^{(k)})\triangleq L(\phi^{(k)})+(\phi-\phi^{(k)})^\intercal \nabla L(\phi^{(k)})+\frac{\omega}{2}(\phi-\phi^{(k)})^\intercal W(\phi-\phi^{(k)}),
\end{equation}
    For $G(\cdot|\cdot)$ defined in Section~\ref{sec:pl2} and $\beta=(\beta_1, ..., \beta_p)^\intercal$, denote
	\begin{equation*}
		\Lambda(\beta)\triangleq \sum_{j=1}^{p} \left(\alpha G(\beta_j|\beta_j^{(k)}) + \lambda\frac{1-\alpha}{2}\beta_j^2\right).
	\end{equation*}
Finally we define the surrogate loss
\begin{equation}\label{eqn:plik3}
	\begin{aligned}
		Q(\phi|\phi^{(k)})&\triangleq  \ell(\phi|\phi^{(k)}) + \Lambda(\beta). 
	\end{aligned}
\end{equation}
A similar result as in \citet{schifano2010majorization} is obtained.
\begin{theorem}\label{thm:mmpg}
    Assume twice differentiable function $F(\phi)$ satisfies (\ref{eqn:omega}) or (\ref{eqn:omega1}), $Q(\phi|\phi^{(k)})$ is given by (\ref{eqn:plik3}), and penalty $p_\lambda(|\cdot|)$ satisfies Assumption~1. Then: 
    $Q(\phi|\phi^{(k)})$ majorizes $F(\phi)$ at $\phi^{(k)}$:
	\begin{equation}\label{eqn:mmp}
	Q(\phi|\phi^{(k)}) \geq F(\phi), \quad 
	Q(\phi^{(k)}|\phi^{(k)}) = F(\phi^{(k)}).
\end{equation}
The inequality holds if $\omega > \varrho$.
	\end{theorem}

Theorem~\ref{thm:mmpg} implies that Algorithm~\ref{alg:alg1} 
provides a nonincreasing sequence $F(\phi^{(k)})$. We analyze Algorithm~\ref{alg:alg1} under different assumptions in the sequel. The following convergence analysis does not cover (\ref{eqn:lqa}) which is not defined when $\theta=0$. See \citet{hunter2005variable} for a modified majorization to avoid such a degenerate case. We first focus on convex optimization problems.
\begin{theorem}\label{thm:mm2}
   Assume twice differentiable convex function $F(\phi)$ satisfies (\ref{eqn:omega}) or (\ref{eqn:omega1}),      $Q(\phi)$ is given by (\ref{eqn:plik3}), the surrogate penalty $G(\beta_j|\beta_j^{(k)})=\lambda|\beta_j|, \lambda > 0$, and the iterates $\phi^{(k)}$ are generated by Algorithm~\ref{alg:alg1}.
    Then
			the loss values $F(\phi^{(k)})$
	are monotonically decreasing and $\lim_{k \to \infty}F(\phi^{(k)}) \to \inf F(\phi)$. 
	Furthermore $\phi^{(k)}$ converges to a minimum point of $F(\phi)$
	if $L(\cdot)$ is either (i) coercive or (ii) bounded below with bounded intercept $\beta_0$.
\end{theorem}

    The bounded intercept in Theorem~\ref{thm:mm2} ensures that penalty term $\Lambda(\beta)$ is coercive for $\phi$. Otherwise $\Lambda(\beta)$ is not coercive for $\phi$ despite that $\Lambda(\beta)$ is coercive for $\beta$ meaning $\norm{\beta} \to \infty$ implies $\Lambda(\beta) \to \infty$. To see this, denote
    \begin{equation*}
		\begin{array}{cc}
				A=(0_{p\times 1} & I_{p\times p}).
	\end{array}
\end{equation*}
Then $A\phi=\beta$. Denote $\tilde{\Lambda}(\phi)\triangleq\Lambda(A\phi)$ and
    let $\beta=(0, ..., 0)^\intercal$, a $p$-dimensional vector, then we have $\phi=(\beta_0, \beta)^\intercal$ and $\tilde{\Lambda}(\phi)= \Lambda(\beta)=0$ by definitions. Thus
    $\Lambda(\beta)$ is a constant as $|\beta_0|\to \infty$ or $\norm{\phi} \to \infty$. In conclusion, as $\norm{\phi} \to \infty$, $\tilde{\Lambda}(\phi) \to \infty$ doesn't always hold. This proves that $\Lambda(\beta)$ is not coercive for $\phi$ if the intercept is not bounded. 
	
The next two theorems are particularly relevant for nonconvex functions. We consider a larger class of surrogate loss $\ell(\phi|\phi^{(0)})$ which covers the QM (\ref{eqn:thm1.38}) as a special case while the surrogate loss function $Q(\phi|\phi^{(k)})$ has the same form as in (\ref{eqn:plik3}).
\begin{theorem}\label{thm:mmpg1}
    Assume that $L(\phi)$ and $\ell(\phi|\phi^{(k)})$ are differentiable functions, $\ell(\phi|\phi^{(k)})$ is jointly continuous in $(\phi, \phi^{(k)})$, $\ell(\phi|\phi^{(k)})$ majorizes $L(\phi)$ at $\phi^{(k)}$, Assumption~\ref{ass:sur} holds, penalty $p_\lambda(|\cdot|)$ satisfies Assumption~\ref{ass:pen}, the surrogate loss $Q(\phi|\phi^{(k)})$ is given by (\ref{eqn:plik3}) where $G(\beta_j|\beta_j^{(k)})$ is defined by (\ref{eqn:nomm}) or (\ref{eqn:lla}). Then every limit point of the iterates generated by Algorithm~\ref{alg:alg1} 
    is a Dini stationary point of the problem minimizing $F(\phi)$.
\end{theorem}

Notice that $\ell(\phi|\phi^{(k)})$ given by (\ref{eqn:thm1.38}) satisfies assumptions of Theorem~\ref{thm:mmpg1}. We can summarize the implications of Theorem~\ref{thm:mmpg1}: applying the QM to the loss functions in the GLM or Table~\ref{tab:tab1}, and applying an appropriate majorization such as (\ref{eqn:nomm}) or (\ref{eqn:lla}) to the penalty functions including the LASSO, SCAD and MCP, then every limit point of the iterates generated by the MM algorithm is a Dini stationary point.
While a Dini stationary point is always a Clarke stationary point, the converse is not valid. Under different conditions, the iterates of the MM algorithm converges to a Clarke stationary point.
\begin{theorem}\label{thm:mm3}
    Assume continuous function $L(\phi)$ is locally Lipschitz continuous, $L(\phi)$ is coercive or bounded below with bounded intercept $\beta_0$, the set of Clarke stationary points $S$ of $F(\phi)$ is a finite set, the surrogate function $\ell(\phi|\phi^{(k)})$ is strongly convex and strictly majorizes $L(\phi)$ at $\phi^{(k)}$, the surrogate loss $Q(\phi|\phi^{(k)})$ is given by (\ref{eqn:plik3}), the surrogate penalty $G(\beta_j|\beta_j^{(k)})$ is convex and majorizes $p_\lambda(|\beta_j|)$
    at $\beta_j^{(k)}$, and the iterates $\phi^{(k)}$ are generated by Algorithm~\ref{alg:alg1}. 
    Then $\phi^{(k)}$ converges to a Clarke stationary point of the problem minimizing $F(\phi)$.
\end{theorem}

Theorem~\ref{thm:mm3} can be applied to a broad loss functions including those in the GLM and Table~\ref{tab:tab1}. These functions are continuously differentiable, thus locally Lipschitz continuous. Construct a surrogate loss $Q(\phi|\phi^{(k)})$ given by (\ref{eqn:plik3}) with $\omega > \varrho$.
Then $\ell(\phi|\phi^{(k)})$ strictly majorizes $L(\phi)$. From $\nabla^2 \ell(\phi|\phi^{(k)})=\omega W$ we know $\nabla^2 \ell(\phi|\phi^{(k)})$ is positive definite. Suppose Assumption~\ref{ass:pen} holds for $p_\lambda(|\cdot|)$, then $G(\phi|\phi^{(k)})$ is convex and majorizes $p_\lambda(|\beta_j|)$ at $\beta_j^{(k)}$  if $G(\phi|\phi^{(k)})$ is given by (\ref{eqn:nomm}) for the LASSO or (\ref{eqn:lla}). This implies some of the assumptions of Theorem~\ref{thm:mm3} have been met. If
additional assumptions in the theorem are valid, then we deduce that Algorithm~\ref{alg:alg1} iterates converge to a Clarke stationary point. Unlike Theorem~\ref{thm:mmpg1}, Theorem~\ref{thm:mm3} doesn't assume equivalence of the first order Dini derivative between the penalized surrogate loss $Q(\phi|\phi^{(0)})$ and objective function $F(\phi)$. When this does hold, as the QM in (\ref{eqn:thm1.38}) and the surrogate penalty function in 
(\ref{eqn:nomm}) or (\ref{eqn:lla}), Theorem~\ref{thm:mmpg1} provides stronger conclusion than Theorem~\ref{thm:mm3}.

\section{Simulation study}
In the simulated examples, we evaluate performance of the proposed robust variable selection methods, i.e., MM for penalized estimation with Algorithm~\ref{alg:alg3}, and compare with the standard non-robust algorithms. The parameter of SCAD and MCP is fixed at $a=3$. This choice works well for the simulation study. The responses in each data set are randomly contaminated with percentage of v=0, 5, 10 and 20. We generate random samples of training/tuning/test data. Training data are used for
model estimation. Tuning samples are used to select optimal penalty parameters with the smallest
loss values and classification errors in regression and classification problems,
respectively. Test data are used to evaluate prediction accuracy except for example 1 whose prediction accuracy is not from the test data. To evaluate variable selection, we compute sensitivity and specificity, where
  sensitivity is the proportion of number of correctly selected predictors       among truly effective predictors,
  and specificity is the proportion of number of correctly non-selected          predictors among truly ineffective predictors. The intercept is not counted as a variable. A good variable selection method should obtain both sensitivity and specificity close to 1.
We repeat Monte Carlo simulation 100 times in the following examples. 
\subsection{Example in regression}
 \indent Example 1:
  Let $y=\beta_0+X^\intercal\beta+\epsilon$, where $\beta_0=1, \beta=(3, 1.5, 0, 0, 2, 0, 0, 0), \epsilon \sim N(0, 3), X\sim N_8(0, \Sigma)$ with              $\Sigma_{ij}=0.5^{|i-j|}$ for $i, j=1, ..., 8$.
 
  Without the intercept, similar data was generated in \citet{fan2001variable}.
  The response variable $y$ is contaminated by randomly multiplying $1,000$ for v = 0, 5, 10 and 20 percent of training (n=200) and tuning data (n=200).        Different estimation methods are evaluated based on the mean squared  prediction error $E[(\hat{y} - \beta_0-X^\intercal
  \beta)^2]$ \citep{hunter2005variable}. We fix $\sigma=10$ in ClossR for which larger $\sigma$ value is more resistant to outliers.  Table~\ref{ex7err} and Table~\ref{ex7nvar}-\ref{ex7spc} in the supplement display the     results for penalized ClossR methods along with standard least squares (LS), LS-based penalized methods and penalized Huber regression proposed by             \citet{yi2017semismooth}. The best prediction results with the associated variable selection are highlighted in bold. 

  The ClossR-based methods have the best   prediction and outperform the HuberLASSO
  in all four settings of outliers. For the LASSO penalty, ClossR has better prediction than Huber but also selects more variables. For the clean data, ClossR-based methods are slightly better or comparable to LS-based methods when the penalty function is the same. If one believes that prediction accuracy from the penalized LS methods  should be superior to the penalized ClossR in uncontaminated cases, the tiny    differences in Table~\ref{ex7err} perhaps can be explained by random
  variation in finite samples and selection of penalty parameters. An interesting research topic is to theoretically compare prediction between penalized LS and penalized ClossR.

  For clean data, all effective variables are selected, and ClossR-based   SCAD or MCP has better selection than LASSO, consistent with LS-based penalty   functions. For contaminated data, all correct variables are selected in robust  estimation, but not in the penalized LS estimation. In addition, ClossR-based   SCAD or MCP determines fewer ineffective variables than LASSO.

For nonconvex optimization, Algorithm~\ref{alg:alg3} can generate different local estimates depending on the initial values. To illustrate, two methods of initial values elaborated in supplement A.1 are compared in Figure~\ref{fig:path} for loss ClossR and nonconvex penalty SCAD with v=20. Prediction from
 boosting-supported backward solution path is more reliable than the forward
 path. The two largest errors from the latter ultimately generates a greater mean prediction error despite a smaller median value.
\subsection{Examples in classification}
The following data generations are from \citet{wu2007robust} and have also been investigated in \citet{wang2018quadratic, wang2018robust} for robust classification.
 These examples include both linear and nonlinear classifiers.

Example 2: Predictor variables $(x_1, x_2)$ are uniformly sampled      from a unit disk $x_1^2+x_2^2 \leq 1$ and $y=1$ if $x_1 \geq x_2$ and -1        otherwise.

Example 3: Predictor variables $(x_1, x_2)$ are uniformly sampled      from a unit disk $x_1^2+x_2^2 \leq 1$ and $y=1$ if $(x_1-x_2)(x_1+x_2) < 0$     and -1 otherwise.

In each example, we also generate 18 noise variables from uniform[-1, 1]. To add outliers, we randomly select v percent of the data and switch their class labels. With the simulated data, the performance of an algorithm can be compared with the optimal Bayes errors. The training/tuning/test sample sizes are $n=200/200/10,000$ in Example 2 and $1,000/1,000/10,000$ in Example 3. To accommodate nonlinear classifiers in Example 3, the predictor space is doubled to $p=40$ including quadratic terms
$x_{i}^2$ for $i=1, 2, ..., 20$. In Example 2 and 3, no-intercept models have better prediction accuracy from logistic-based and nonconvex loss-based variable selection methods. Since the data are not generated from logistic regression models, estimation from a logistic regression model is not a clear-cut winner  even for the clean data. We fix Closs $\sigma=0.9$, Gloss $\sigma=1.1$ and Qloss $\sigma=0.2$ in the estimation. 

The results for Example 2 are demonstrated in Table~\ref{tab:ex1err} and Table~\ref{tab:ex1nvar} in the supplement. It is clear that the performance of logistic-based methods relies on the cleanliness of the data, and is not robust with outliers. On the other hand, the nonconvex loss-based variable selection methods are better resistant to outliers. In particular, with data contaminated by v=0, 5, 10 and 20 percentage, the best prediction results 0.0047(QlossSCAD), 0.0548(GlossSCAD), 0.1055(GlossSCAD), 0.2067(QlossSCAD) are the
closest to the corresponding Bayes errors. The corresponding variable selections in highlighted numbers in Table~\ref{tab:ex1nvar} almost perfectly match the Bayes optimal classifier. Since all effective variables are correctly selected, specificity is not reported in lieu of supplement Table~\ref{tab:ex1nvar}.
It can be interesting to compare the LASSO and boosting due to its similarity \citep{Efro:lars:2004}. The same random data was generated in \citet{wang2018quadratic}, where the robust boosting algorithm provided similar prediction results and slightly more sparse variable selection. For instance, the boosting errors from the Gloss are 0.0108, 0.0628, 0.1178, 0.2217 for outlier percentage v=0, 5, 10, 20, respectively \citep{wang2018quadratic}, while the corresponding QlossLASSO errors are 0.0154,
0.0672, 0.1187, 0.2223 in Table~\ref{tab:ex1err}. As the percentage of outliers changes, the average number of variables selected in the boosting algorithm ranges from 2.5 to 2.8 \citep{wang2018quadratic} while the QlossLASSO ranges from 4.1 to 4.9 in supplement Table~\ref{tab:ex1nvar}.  

The results for Example 3 are summarized in the supplement, and similar conclusions like Example 2 can be drawn. In all outlier scenarios, the nonconvex loss-based methods triumph over the logistic-based methods on prediction. In particular, the best average test errors are 0.0016 (ClossMCP), 0.0520 (QlossSCAD), 0.1018 (QlossSCAD) and 0.2019 (QlossSCAD) for percentage of outliers v=0, 5, 10 and 20, respectively. In addition, the corresponding variable selections in featured
numbers in Table~\ref{tab:ex2nvar} resemble the Bayes optimal classifier. Except for v=0, all variables are more accurately identified by the nonconvex loss-based methods than the logistic-based methods. 
Both $x_1^2$ and $x_2^2$ are selected in all estimation methods in each simulation      run. Supplement Table~\ref{tab:ex2_var} counts variable selection for $x_1$ or $x_2$ in 100 simulations.       Across all methods, the average specificity excluding $x_1, x_1^2, x_2, x_2^2$ is $\geq 0.98$.  Hence, the details are unreported. In conclusion, all methods correctly select $x_1^2$ and      $x_2^2$, and the remaining variables including $x_1$ and $x_2$     are typically eliminated by the variable
 selection procedures.
\section{Predicting breast cancer clinical status} 
\citet{maqc2010microarray} used gene expressions to predict breast cancer clinical status with 164 estrogen receptor positive cases and 114 negative cases. The same data set has been evaluated for a variety of robust classification algorithms \citep{li2018boosting, wang2018quadratic, wang2018robust}. Applying MM for penalized estimation, the analysis described below is reproduced in a vignette distributed with the \texttt{R} package \texttt{mpath}. Following the same pre-selection procedure, the data set is pre-screened to keep 3000 most significant genes from 22,282 genes with two-sample t-test. These genes are then standardized. The data set is randomly split
to training data and test data, and the training data contains 50/50 positive/negative cases and the test data has 114/64 positive/negative cases. The positive/negative status in the training data is randomly flipped with percentage v=0, 5, 10, 15. The parameter of SCAD and MCP is fixed at $a=3.7$ and $a=12$, respectively. The predictive models are aided with a five-fold cross-validation to select the optimal penalty parameters with the smallest classification errors. The analysis is conducted with Closs $\sigma=0.9$, Gloss $\sigma=1.01$ and Qloss $\sigma=0.5$ \citep{wang2018quadratic}. This process is
repeated 100 times and the results are summarized in Table~\ref{tab:cancererr} and supplement Table~\ref{tab:cancernvar}. 

Prediction accuracy is more similar in the clean data than those contaminated, while the logistic-based methods suffer from outliers. The best prediction with the Closs-based methods are 0.0818, 0.0839, 0.0910, 0.1143 for v=0, 5, 10, 15, respectively. This is the best prediction for this data set in the robust prediction literature. See \citet{li2018boosting, wang2018quadratic, wang2018robust} and the
references therein. With more data contamination, more variables are selected. Loss function
and penalty method also contribute to different variable selection results.
In addition, sparse variable selection is maintained even with outliers. For instance, with 15\% outliers, ClossSCAD not only has more accurate prediction than LogisSCAD (0.1143 vs 0.1566), but also selected only 2.8 genes on average, compared with 19.1 genes by LogisSCAD.

\section{Discussion}
Outliers can substantially cause bias in traditional estimation methods and produce inaccurate variable selection problem. This problem has been tackled by robust estimation, which has a long history in statistics methodology research and applications. The previous methods are largely based on convex-based loss functions, which may not be sufficient to guard the impact of outliers. Adding to this is the variable selection challenge in predictive modeling, specifically for the high-dimensional data. 

In this article, we propose penalized nonconvex loss functions for several popular penalty functions. We have illustrated that the proposed methods can improve estimation and variable selection compared to conventional estimation and robust estimation. The estimation problem is decoupled into an iterative MM algorithm with each iteration conveniently conducted by existing optimization algorithms. The QM embedded in the MM algorithm is particularly appealing since the induced penalized least
squares problem has been well studied. For tuning parameter identification, it is a typical practice to construct a solution path along with penalty parameters. A popular procedure is to start with a penalty parameter $\lambda_{\max}$ such that all predictor coefficients are zero. As is well-known in nonconvex optimization problems, different starting values may lead to different solutions if local minimizers are achieved. We propose an alternative yet effective approach to compute a solution path. The so-called backward path generates starting values utilizing the boosting algorithm and find solutions backwards towards zero coefficients. 

This article evaluates optimality conditions for both convex and nonconvex problems in the context of penalized estimation. As expected, we could only partially characterize the optimality conditions for nonconvex problems. The link between the original optimization problem and the surrogate optimization problem, nevertheless, provides a useful utility to compute $\lambda_{\max}$. The convergence analysis provides new results for both convex and nonconvex problems. For convex problems, the iterated estimates of the MM algorithm with QM converges to a minimum point. For nonconvex problems, the iterates of the MM algorithm with QM converges to a local stationary point in the framework of the Dini or Clarke derivative. 

Future direction of research includes expanded penalty functions including those not locally Lipschitz continuous. For instance, the bridge penalty for $a \in (0, 1)$. Since the generalized derivatives like Dini and Clarke derivative are both restricted to locally Lipschitz continuous functions, additional mathematical tools are required to assess properties of the penalized loss functions.
\bibliographystyle{apalike}
\bibliography{../../wangres}

\begin{figure}[!htbp]
  \begin{center}
      \includegraphics[scale=0.3]{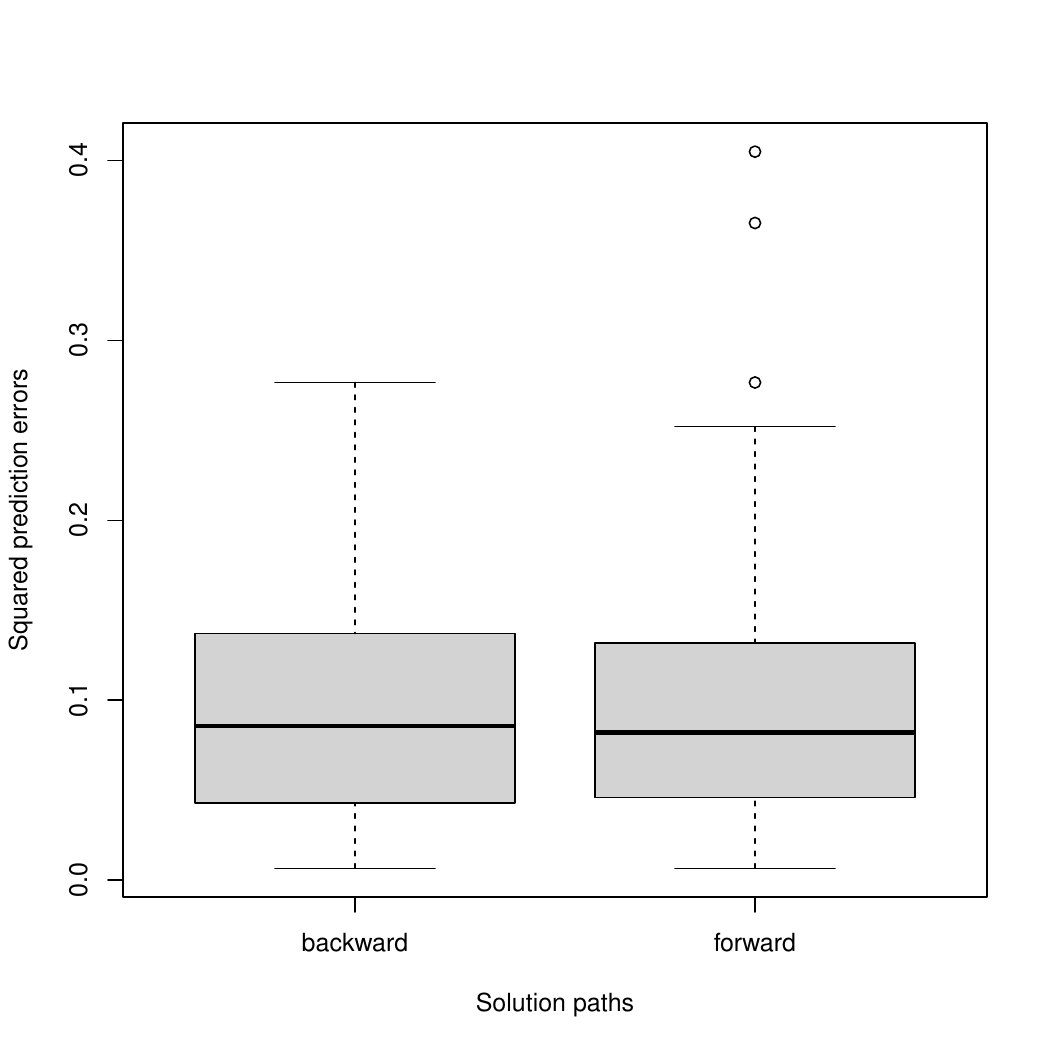}
    \caption{Prediction of ClossRSCAD from different solution paths in Example 1 and v=20}
    \label{fig:path}
  \end{center}
\end{figure}

\begin{table}[!hbtp]
	\begin{center}
\begin{tabular}{ lc}
	\hline\hline
	Nonconvex loss & $\Gamma(u)$\\
\hline
	ClossR & $1 - \exp(\frac{-u^2}{2\sigma^2}), \sigma > 0$\\
	Closs  & $c\left(1 - \exp(\frac{-(1-u)^2}{2\sigma^2})\right), c=\frac{1}{1 - \exp(\frac{-1}{2\sigma^2})}, \sigma > 0 $\\ 
Gloss & $\frac{2^\sigma}{\left(1+\exp(u)\right)^\sigma}, \sigma > 1$\\ 
Qloss & $ 2\left(1- \Phi(\frac{u}{\sigma})\right), \Phi(\frac{u}{\sigma})=\int_{-\infty}^{\frac{u}{\sigma}}\frac{1}{\sqrt{2\pi}}\exp(\frac{-x^2}{2})dx, \sigma > 0$\\
\hline
\hline
\end{tabular}
	\end{center}
\caption{Nonconvex loss function}
\label{tab:tab1} 
\end{table}

\begin{table}[!tbp]
    \caption{Mean squared prediction errors and standard deviations (in parentheses) for v percentage of outliers in Example 1. Numbers $> 1\mathrm{e}+4$ are suppressed.
\label{ex7err}}
	\begin{center}
		\begin{tabular}{lllll}
			\hline\hline
			\multicolumn{1}{l}{Method}&\multicolumn{1}{c}{v=0}&\multicolumn{1}{c}{v=5}&\multicolumn{1}{c}{v=10}&\multicolumn{1}{c}{v=20}\tabularnewline
\hline
            LS&0.1358(0.0629)&&&\tabularnewline
			LSLASSO&0.0910(0.0479)&&&\tabularnewline
            LSSCAD&0.0702(0.0454)&&&\tabularnewline
			LSMCP& 0.0702(0.0454)&&&\tabularnewline
			HuberLASSO&0.1232(0.0718)&0.1137(0.0737)&0.1198(0.0708)&0.2056(0.1261)\tabularnewline
            ClossRLASSO&0.0905(0.0480)&0.0966(0.0507)&0.1023(0.0561)&0.1216(0.0643)\tabularnewline
            ClossRSCAD&0.0718(0.0443)&0.0787(0.0505)&0.0834(0.0689)&\textbf{0.0917}(0.0574)\tabularnewline
            ClossRMCP&\textbf{0.0699}(0.0468)&\textbf{0.0749}(0.0500)&\textbf{0.0787}(0.0594)&0.0953(0.0656)\tabularnewline
\hline
\hline
\end{tabular}\end{center}

\end{table}

\begin{table}[!tbp]
    \caption{Mean test errors and standard deviations (in parentheses) for v pencentage of outliers in Example 2.}
    \label{tab:ex1err}
	\begin{center}
		\begin{tabular}{lllll}
			\hline\hline
			\multicolumn{1}{l}{Method}&\multicolumn{1}{c}{v=0}&\multicolumn{1}{c}{v=5}&\multicolumn{1}{c}{v=10}&\multicolumn{1}{c}{v=20}\tabularnewline
\hline
            Bayes &0&0.05&0.1&0.2\tabularnewline
            Logis&0.0464(0.0127) &0.1272(0.0154)&0.1842(0.0146) &0.2908(0.0158)\tabularnewline
            LogisLASSO&0.0142(0.0097) &0.0728(0.0147)&0.1266(0.0178)&0.2324(0.0215)\tabularnewline
            LogisSCAD&0.0048(0.0048)&0.0619(0.0117)&0.1215(0.0142)&0.2286(0.0195)\tabularnewline
            LogisMCP&0.0072(0.0065)&0.0677(0.0124)&0.1219(0.0150)&0.2356(0.0185)\tabularnewline
            ClossLASSO&0.0242(0.0163) &0.0761(0.0182)&0.1274(0.0167)&0.2292(0.0170)\tabularnewline
            ClossSCAD&0.0181(0.0119) &0.0685(0.0127)&0.1186(0.0135)&0.2187(0.0129)\tabularnewline
            ClossMCP&0.0181(0.0119) &0.0685(0.0127)&0.1192(0.0139)&0.2197(0.0134)\tabularnewline
            GlossLASSO&0.0141(0.0098) &0.0662(0.0103)&0.1177(0.0120)&0.2224(0.0150)\tabularnewline
            GlossSCAD&0.0047(0.0044) &\textbf{0.0548}(0.0044)&\textbf{0.1055}(0.0054)&0.2103(0.0131)\tabularnewline
            GlossMCP&0.0050(0.0048) &0.0554(0.0050)&0.1072(0.0070)&0.2086(0.0107)\tabularnewline
            QlossLASSO&0.0154(0.0103) &0.0672(0.0103)&0.1187(0.0121)&0.2223(0.0146)\tabularnewline
            QlossSCAD&\textbf{0.0047}(0.0043) &0.0549(0.0045)&0.1056(0.0053)&\textbf{0.2067}(0.0065)\tabularnewline
            QlossMCP&0.0049(0.0045) &0.0554(0.0048)&0.1056(0.0054)&0.2070(0.0066)\tabularnewline
\hline
\hline
\end{tabular}\end{center}

\end{table}

\begin{table}[!tbp]
    \caption{Mean test errors and standard deviations (in parentheses) for v pencentage of outliers in the breast cancer data.}\label{tab:cancererr}
	\begin{center}
		\begin{tabular}{lllll}
			\hline\hline
			\multicolumn{1}{l}{Method}&\multicolumn{1}{c}{v=0}&\multicolumn{1}{c}{v=5}&\multicolumn{1}{c}{v=10}&\multicolumn{1}{c}{v=15}\tabularnewline
			\hline
			~~LogisLASSO&0.0851(0.0143) & 0.1058(0.0262)  & 0.1236(0.0356)   &0.1567(0.0406) \tabularnewline
			~~LogisSCAD&0.0846(0.0127) & 0.1051(0.0264)  & 0.1289(0.0359)   &0.1566(0.0409) \tabularnewline
			~~LogisMCP&0.0844(0.0132) & 0.1032(0.0268)  & 0.1289(0.0397)   &0.1608(0.0443) \tabularnewline
            ~~ClossLASSO&0.0818(0.0145) & 0.0860(0.0169)  & 0.0924(0.0256)   &0.1171(0.0461) \tabularnewline
            ~~ClossSCAD&0.0819(0.0128)  & 0.0859(0.0243)  & 0.0924(0.0334)   &\textbf{0.1143}(0.0519) \tabularnewline
            ~~ClossMCP  &\textbf{0.0805}(0.0115) & \textbf{0.0839}(0.0261)  & \textbf{0.0910}(0.0358)   &0.1150(0.0598) \tabularnewline
			~~GlossLASSO&0.0833(0.0138) & 0.0858(0.0176)  & 0.0974(0.0293)   &0.1261(0.0467) \tabularnewline
            ~~GlossSCAD&0.0836(0.0128)  & 0.0878(0.0229)  & 0.0972(0.0432)  & 0.1330(0.0723) \tabularnewline
			~~GlossMCP  &0.0820(0.0135) & 0.0854(0.0239)  & 0.0934(0.0344)   &0.1312(0.0597) \tabularnewline
			~~QlossLASSO&0.0822(0.0138) & 0.0849(0.0160)  & 0.0974(0.0281)   &0.1249(0.0447) \tabularnewline
			~~QlossSCAD  &0.0833(0.0129) & 0.0893(0.0257)  & 0.0974(0.0435)   &0.1264(0.0657) \tabularnewline
            ~~QlossMCP&0.0820(0.0132)  & 0.0848(0.0214)  & 0.0935(0.0342)  & 0.1271(0.0584) \tabularnewline
\hline                 
\end{tabular}\end{center}
\end{table}



\clearpage
\setcounter{page}{0}
    \pagenumbering{arabic}
    \setcounter{page}{1}
\begin{center}
    \textbf{MM for Penalized Estimation}\\
Supplementary Information\\
Zhu Wang\\
    \email{zhuwang@gmail.com}
\end{center}
\begin{appendices}
    \section{Technical details of Algorithm~\ref{alg:alg3}}
    \subsection{Solution paths}\label{s:tuning}
To optimize $F(\phi)$, tuning parameters such as $\lambda$ can be chosen by cross-validation or other data driven methods from some prespecified values. We make use of $\lambda_{\max}$ in (\ref{eqn:opt3}) or equivalently $\tilde\lambda_{\max}$ in (\ref{eqn:thm3.4}). The latter can be handy with existing computer codes for penalized least squares estimation. We compute the solutions along a regularized path for a sequence of $\lambda$. A forward path begins with all initial non-intercept
    parameters zero for a fixed $\alpha$ value, then compute solutions for a decreasing sequence of $\lambda$, $\lambda_{k}, k=1, ..., K$, ranging from $\lambda_{\max}$ to $\lambda_{\min}=\epsilon\lambda_{\max}$ for a small number $0<\epsilon<1$ \citep{Frie:regu:2010}. 
    For nonconvex estimation in Section~\ref{sec:robust}, Algorithm~\ref{alg:alg3} typically seeks a local optimization
solution. It is well-known that there could be potentially different estimates
depending on the initial values. In practice, different initial values can be utilized to obtain the optimal result. 
    This article emphasizes
boosting as an innovation for initial value estimation, coupled with an alternative backward path, an increasing sequence of $\lambda_{k}, k=K, ..., 1$ from $\lambda_{\min}$ to $\lambda_{\max}$. This
would provide different initial values for nonconvex optimization.
    Most importantly, the numerical results in the simulation and data analysis warrant this choice.
\begin{table}[htbp!]
	\begin{center}
\begin{tabular}{ lc}
	\hline\hline
	Nonconvex loss & $\lambda_{\max}$\\
\hline
	ClossR           & $\max_{1 \leq j \leq p} \frac{1}{n\alpha \sigma^2}\left|\sum_{i=1}^n x_{ij} (y_i-\beta_0) \exp\left(-\frac{(y_i-\beta_0)^2}{2\sigma^2}\right) \right| $\\
	Closs             & $\max_{1 \leq j \leq p} \frac{c}{n\alpha \sigma ^2}\left|\sum_{i=1}^n x_{ij}(y_i-\beta_0)\exp\left(-\frac{(1-y_i\beta_0)^2}{2\sigma^2}\right)\right|$\\
	Gloss             & $\max_{1 \leq j \leq p} \frac{\sigma2^\sigma}{n\alpha} \left|\sum_{i=1}^n x_{ij} y_i \exp(y_i \beta_0)(1+\exp(y_i \beta_0))^{-\sigma-1}\right|$\\
	Qloss             & $\max_{1 \leq j \leq p} \frac{\sqrt{2}}{n \alpha \sqrt{\pi} \sigma} \exp(-\frac{\beta_0^2}{2\sigma^2})\left| \sum_{i=1}^n x_{ij} y_i \right|$\\
\hline
\hline
\end{tabular}
	\end{center}
	\caption{The $\lambda_{\max}$ values}
\label{tab:tab3} 
\end{table}

    \subsection{Update on active set}
 To speed up computations, we may implement the active set strategy in Algorithm~\ref{alg:alg3}. First, we determine the active set, a set of variables with non-zero coefficients, from the initial value $\phi$ in step 1. Run step 2 to 5 with variables in the active set. Then run step 4 once with all variables in the   full set to update the active set. If there is no change, the computation finishes. Otherwise, repeat the above procedure. Second, we employ the active set    internally within step
    4 when
 minimizing penalized least squares. We begin with an initial active set containing all variables, cycle through all coefficients in the active set with        coordinate descent algorithm until convergence, then cycle through the full set with all variables, but only once. This generates a new active set. Compared   with the previous active set, if no change, then step 4 converges. Otherwise, we repeat the above process until convergence. 
The first two procedures invoke the full set so that variable selection is not predominated or trapped by the initial estimation including boosting.   Indeed, if all the initial parameters are zero (forward path) in step 1, the above full set procedures still apply. That is, the variable selection is not     completely relying on step 1.
    Third, for large data problems,   the active set principle can be aggressively followed through to develop a solution path for a desired $\lambda$ sequence. For an increasing sequence of       $\lambda$, starting from a small $\lambda_K(=\lambda_{\min})$ value, Algorithm~\ref{alg:alg3} generates estimates and an active set. For $\lambda_{K-1} (>     \lambda_{K})$, only the coefficients of the active set are cycled through while the coefficients of the non-active set remain zero. The process is repeated    until $\lambda_1(=\lambda_{\max})$ is taken.
    
Employing active set can substantially reduce computing burden          especially for large data problems as in Section 6. The results, however, are   not always preferred compared to the conventional non-active set, or full set strategy. Therefore, we focus on         results from the non-active set strategy in the numerical studies.
    \section{Generalized derivatives}\label{app:gen}
Generalized derivatives for nonsmooth convex and nonsmooth nonconvex functions have been developed \citep{nesterov2004introductory, clarke2013functional}. Throughout this paper, we only consider $f: \mathbb{R}^m \to \mathbb{R}$, a locally Lipschitz continuous function. 
    It is said to be Lipschitz near $x$ if, for some neighborhood $V_x$ of $x$ and some constant $K$ such that $|f(x)-f(y)|\leq K\norm{x-y} \ \forall x, y \in V_x$. And $f$ is called locally Lipschitz continuous  
    on $\mathbb{R}^m$ if it is Lipschitz near $x$ for every $x \in \mathbb{R}^{m}$. 
    If for all $y$, $f(y)-f(x) \geq \langle g, y-x \rangle$, then $g$ is called a subgradient of $f$ at $x$. The ordinary directional derivative at $x$ in the direction $\varepsilon$ is defined by:
\begin{equation*}
f'(x; \varepsilon)\triangleq\lim\limits_{\tau\to 0+}\frac{f(x+\tau \varepsilon)-f(x)}{\tau}.
\end{equation*}
    The lower directional Dini derivative of $f$ at $x$ in the direction $\varepsilon$ is defined below:
\begin{equation*}
f'_D(x; \varepsilon)\triangleq\liminf\limits_{\tau\to 0+}\frac{f(x+\tau \varepsilon)-f(x)}{\tau}.
\end{equation*}
    The Clarke directional derivative of $f$ at $x$ in the direction $\varepsilon$ is defined below:
\begin{equation*}
f^o(x; \varepsilon)\triangleq\limsup\limits_{y \to x, \tau\to 0+}\frac{f(y+\tau \varepsilon)-f(y)}{\tau}.
\end{equation*}
The Clarke subdifferential of $f$ at $x$ is the set \citep[10.3]{clarke2013functional} 
    \begin{equation*}
        \partial_C f(x)=\{v \in \mathbb{R}^m: f^o(x, d) \geq v^\intercal d \text{ for all } d \in \mathbb{R}^m\}.
\end{equation*}
We say that $f$ is regular at $x$ provided $f$ is Lipschitz near $x$ and
admits directional derivatives $f'(x ; \varepsilon)$ satisfying $f^o(x ; \varepsilon) = f'(x ; \varepsilon) \ \forall \varepsilon$.
    If $f$ is differentiable at $x$, then all three directional derivatives at $x$ are the same. However, unlike the directional derivative $f'(x; \varepsilon)$, the Dini and Clarke derivatives always exist.
    The point $x$ is a stationary point of $f(\cdot)$ if $f'(x; \varepsilon) \geq 0, \forall\ \varepsilon \in \mathbb{R}^m$. 
The point $x$ is a Dini stationary point of $f(\cdot)$ if $f'_D(x; \varepsilon) \geq 0, \forall\ \varepsilon \in \mathbb{R}^m$. 
The point $x$ is a Clarke stationary point of $f(\cdot)$ if $f^o(x; \varepsilon) \geq 0, \forall\ \varepsilon \in \mathbb{R}^m$. 
When a directional derivative exists, a stationary point is always a Dini stationary point but not vice versa. 
A Dini stationary point is always a Clarke stationary
point but not vice versa since the following relationships hold for all $x$ and $\varepsilon$: 
\begin{equation*}
	\begin{aligned}
		f'_D(x; \varepsilon)&=\liminf\limits_{\tau\to 0+}\frac{f(x+\tau \varepsilon)-f(x)}{\tau} \\
		 &\leq \limsup\limits_{\tau\to 0+}\frac{f(x+\tau \varepsilon)-f(x)}{\tau} \\
		 &\leq \limsup\limits_{y \to x, \tau\to 0+}\frac{f(y+\tau \varepsilon)-f(y)}{\tau}\\
		 &=f^o(x; \varepsilon).
	\end{aligned}
\end{equation*}
We say that a vector $x\in \mathbb{R}^m$ is a limit point of a sequence $\{x_n\}$ in $\mathbb{R}^m$ if there exists a subsequence of $\{x_n\}$ that converges to $x$. 

    \section{Proofs}\label{app:pro}
\begin{lemma}\label{lem:pen2}
    If Assumption~\ref{ass:pen2} holds, then:
    \begin{equation}\label{eqn:lem1_0}
        \partial_C p_\lambda(0)\supseteq [-\zeta\lambda, \zeta\lambda],
    \end{equation}
    where $\partial_C p_\lambda(0)$ is the Clarke subdifferential of $p_\lambda(|\theta|)$ at $\theta=0$.
Equality holds if $p_\lambda(|\theta|)$ is also regular at $\theta=0$.
\end{lemma}
    \begin{remark}
        Lemma~\ref{lem:pen2} can be applied to the LASSO penalty with equality in (\ref{eqn:lem1_0}) since the penalty is convex, thus a regular function \citep[10.8]{clarke2013functional}. A regular function is either smooth everywhere or else has a corner of convex type \citep[p. 200]{clarke2013functional}. Under Assumption~\ref{ass:pen} or Assumption~\ref{ass:pen2}, $p_\lambda(|\theta|)$ is not differentiable at $\theta=0$, and indeed $0$ is a corner. Therefore $p_\lambda(|\theta|)$ is
        regular if and only if $p_\lambda(|\theta|)$ is convex with $0$ corner, i.e., the LASSO penalty. On the other hand, penalty functions SCAD and 
        MCP are not regular as they are folded concave functions on $\mathbb{R}$ with a corner at the origin. Hence (\ref{eqn:lem1_0}) may not hold for these penalty functions.
    \end{remark}
        \noindent{\textbf{Proof of Lemma~\ref{lem:pen2}}}

Let $p^o_\lambda(0; \varepsilon)$ denote the Clarke directional derivative at $0$ in the direction $\varepsilon$, and $p'_\lambda(0; \varepsilon)$ denote the ordinary directional derivative. Their relationship and Assumption~\ref{ass:pen2} lead to 
\begin{equation*}
    p^o_\lambda(0; \varepsilon) \geq p'_\lambda(0; \varepsilon)=\zeta\lambda |\varepsilon|. 
\end{equation*}
Let $\tau \in [-\zeta\lambda, \zeta\lambda]$, i.e. $\zeta\lambda \geq |\tau|$, then we get
\begin{equation*}\label{eqn:lem1_1}
    p^o_\lambda(0; \varepsilon) \geq \langle{\tau, \varepsilon\rangle}=\tau\varepsilon \ \forall \varepsilon.
\end{equation*}
By definition \citep[10.3]{clarke2013functional}, $\tau \in \partial_C p_\lambda(0)$. Thus we have shown 
\begin{equation}\label{eqn:lem1_1.1}
    \partial_Cp_\lambda(0) \supseteq [-\zeta\lambda, \zeta\lambda].
\end{equation}
If $p_\lambda(\cdot)$ is also regular at $0$, by its definition and Assumption~\ref{ass:pen2}, we then have 
\begin{equation}\label{eqn:lem1_2}
    p^o_\lambda(0; \varepsilon) = p'_\lambda(0; \varepsilon)=\zeta\lambda |\varepsilon|. 
\end{equation}
Let $\tau \in \partial_Cp_\lambda(0)$, then according to the definition \citep[10.3]{clarke2013functional}, a necessary and sufficient condition is
\begin{equation*}\label{eqn:lem1_3}
    p^o_\lambda(0; \varepsilon) \geq \langle \tau, \varepsilon \rangle=\tau \varepsilon \ \forall \varepsilon.
\end{equation*}
Along with (\ref{eqn:lem1_2}), then $\zeta\lambda|\varepsilon| \geq \tau\varepsilon$, which leads to $\tau \in [-\zeta\lambda, \zeta\lambda]$. Hence $\partial_Cp_\lambda(0) \subseteq [- \zeta\lambda, \zeta\lambda]$. Combining with (\ref{eqn:lem1_1.1}), we have $\partial_Cp_\lambda(0) =   [-\zeta\lambda, \zeta\lambda]$.
\qed

\noindent{\textbf{Proof of Theorem~\ref{thm:opt}}}

    Part (a):
    First, $F(\phi)$ is locally Lipschitz continuous as $F(\phi)$ is a sum of locally Lipschitz continuous functions. 
    Since the first summand $L(\phi)$ can be nonconvex and the second summand is nonsmooth, we invoke the Clarke subdifferential. Suppose $\phi^\ast$ is a local minimizer of $F(\phi)$, then $0 \in \partial_C F(\phi^\ast)$, where $F(\phi^\ast)$ is the Clarke subdifferential \cite[10.7(a)]{clarke2013functional}.
We have $L(\phi)$ continuously differentiable, $\beta_j^2$ continuously differentiable and $p_\lambda(|\beta_j|)$ Lipschitz continuous near $\beta_j$, then the calculus of Clarke subdifferential holds \citep[10.16]{clarke2013functional}: 
\begin{equation}\label{eqn:clarke0}
    \partial_C F(\phi)=\left\{\nabla L(\phi)\right\}+\partial_C\sum_{j=1}^p\alpha p_\lambda(|\beta_j|)+\left\{\lambda(1-\alpha)(0, \beta)^\intercal\right\}.
\end{equation}
Once $\phi^\ast$ is plugged into (\ref{eqn:clarke0}), we then obtain
\begin{equation*}\label{eqn:clarke01}
    0 \in \partial_C F(\phi^\ast)=\left\{\nabla L(\phi^\ast)\right\}+\partial_C\sum_{j=1}^p\alpha p_\lambda(|\beta_j^\ast|)+\left\{\lambda(1-\alpha)(0, \beta^\ast)^\intercal\right\}.
\end{equation*}
If $\beta_j^\ast \neq 0$, $p_\lambda(|\beta_j^\ast|)$ is continuously differentiable by Assumption~\ref{ass:pen}, thus $\partial_C p_\lambda(|\beta_j^\ast|)=\{\text{sign}(\beta_j^\ast)p'_\lambda(\beta_j^\ast)\}$ \citep[10.8]{clarke2013functional}. Hence (\ref{eqn:opt1}) follows. Otherwise we obtain (\ref{eqn:opt1.2}). 

Part (b): First, part (a) holds since Assumption~\ref{ass:pen2} contains Assumption~\ref{ass:pen}. In particular, since (\ref{eqn:opt1.2}) holds, it is left to compute $\partial_C p_\lambda(0)$ utilized in (\ref{eqn:opt1.2}).
For the first statement, Lemma~\ref{lem:pen2} implies 
$\partial_C p_\lambda(0)\supseteq [-\zeta\lambda, \zeta\lambda]$. 
Thus (\ref{eqn:opt1.2}) induces conditions including but potentially not limited to (\ref{eqn:opt2}). In other words, it is possible (\ref{eqn:opt2}) is an incomplete optimality condition. For the second statement, we provide two proofs. (i) Lemma~\ref{lem:pen2} implies
 $\partial_C p_\lambda(0)= [-\zeta\lambda, \zeta\lambda]$. Hence (\ref{eqn:opt1.2}) is equivalent to (\ref{eqn:opt2}), which becomes the complete optimality condition. (ii) As argued before, if $p_\lambda(|\theta|)$ satisfies Assumption~\ref{ass:pen2} and is also regular at $0$, then $p_\lambda(|\theta|)$ is the LASSO penalty. 
For the continuous convex LASSO penalty, the Clarke subdifferential is equivalent to the ordinary subdifferential \cite[10.8]{clarke2013functional}. Therefore $\partial_Cp_\lambda(\cdot)=\partial p_\lambda(\cdot)=[-1, 1]$. From (\ref{eqn:opt1.2}) we again obtain complete optimal condition (\ref{eqn:opt2}). 
\qed

\noindent{\textbf{Proof of Theorem~\ref{thm:opt1.1}}}

We know $\phi^\ast$ is a global minimizer if and only if $0 \in \partial F(\phi^\ast)$ \citep{nesterov2004introductory}. Since $F(\phi^\ast)$ is a sum of convex functions, we have
\begin{equation*}
    \begin{aligned}
        \partial F(\phi^\ast)&=\partial L(\phi^\ast)+\partial\sum_{j=1}^p\alpha p_\lambda(|\beta_j^\ast|)+\partial\sum_{j=1}^p\lambda\frac{(1-\alpha)}{2}(\beta_j^\ast)^2\\
        &=\left\{\nabla L(\phi^\ast)\right\}+\partial\sum_{j=1}^p\alpha p_\lambda(|\beta_j^\ast|)+\left\{\lambda(1-\alpha)(0, \beta^\ast)^\intercal\right\},
    \end{aligned}
\end{equation*}
where the last equality is obtained since $L(\phi)$ and $\beta^2$ are differentiable convex functions. Deduce from this that (\ref{eqn:opt2.1}) and (\ref{eqn:opt2.1_1}) are valid by simple calculations.
\qed
	
    \noindent{\textbf{Proof of Lemma~\ref{lem:mm}}}

    Suppose $u^\ast$ is a local (global) minimizer of $\Gamma(u)$. Using (\ref{eqn:mm0}), the following holds:
    \begin{equation*}
        \gamma(u|u^\ast) \geq \Gamma(u) \geq \Gamma(u^\ast)=\gamma(u^\ast|u^\ast),
    \end{equation*}
    where the second inequality holds locally or globally depending on whether $u^\ast$ is a local or global minimizer. Thus $u^\ast$ is a local (global) minimizer of $\gamma(u|u^\ast)$.
\qed
	
    \noindent{\textbf{Proof of Lemma~\ref{lem:mm2}}}

Suppose $\Gamma(u)$ and $\gamma(u|z)$ are differentiable convex functions. Then $u^\ast$ is a global minimizer of $\gamma(u|u^\ast)$ if and only if 
$\nabla\gamma(u^\ast|u^\ast)=0$ by the optimality condition. Thus $\nabla\Gamma(u^\ast)=\nabla\gamma(u^\ast|u^\ast)=0$ by assumption. This happens if and only if $u^\ast$ is a global minimizer of $\Gamma(u)$ since $\Gamma(u)$ is convex.
\qed
	
    \noindent{\textbf{Proof of Theorem~\ref{thm:clarkeequ}}}

    Part(a): If $\phi^\ast$ is a local minimizer of $F(\phi)$, by Lemma~\ref{lem:mm}, $\phi^\ast$ is also a local minimizer of $Q(\phi|\phi^\ast)$. Therefore $0 \in \partial_C Q(\phi^\ast|\phi^\ast)$. As in the proof of Theorem~\ref{thm:opt} (a), we have
\begin{equation*}\label{eqn:clarke3}
    \begin{aligned}
        \partial_C Q(\phi^\ast|\phi^\ast)&=\left\{\nabla \ell(\phi^\ast|\phi^\ast)\right\}+\partial_C\sum_{j=1}^p\alpha p_\lambda(|\beta_j^\ast|)+\left\{\lambda(1-\alpha)(0, \beta^\ast)^\intercal\right\}\\
        &=\left\{\nabla L(\phi^\ast)\right\}+\partial_C\sum_{j=1}^p\alpha p_\lambda(|\beta_j^\ast|)+\left\{\lambda(1-\alpha)(0, \beta^\ast)^\intercal\right\}\\
        &=\partial_C F(\phi^\ast),
    \end{aligned}
\end{equation*}
where the second equality is obtained by Assumption~\ref{ass:sur}.

Part (b): The proof follows the same arguments as in Part (b) of Theorem~\ref{thm:opt}.
\qed

	\noindent{\textbf{Proof of Theorem~\ref{thm:mmopt2}}}

    The convexity assumption leads to
\begin{equation}\label{eqn:clarke3.1}
    \begin{aligned}
        \partial Q(\phi^\ast|\phi^\ast)
        &=\partial \ell(\phi^\ast|\phi^\ast)+\partial\sum_{j=1}^p\alpha p_\lambda(|\beta_j^\ast|)+\partial\sum_{j=1}^p\lambda\frac{(1-\alpha)}{2}(\beta_j^\ast)^2\\
        &=\left\{\nabla \ell(\phi^\ast|\phi^\ast)\right\}+\partial\sum_{j=1}^p\alpha p_\lambda(|\beta_j^\ast|)+\left\{\lambda(1-\alpha)(0, \beta^\ast)^\intercal\right\}\\
        &=\left\{\nabla L(\phi^\ast)\right\}+\partial\sum_{j=1}^p\alpha p_\lambda(|\beta_j^\ast|)+\left\{\lambda(1-\alpha)(0, \beta^\ast)^\intercal\right\}\\
        &=\partial F(\phi^\ast),
    \end{aligned}
\end{equation}
where the third equality is obtained by Assumption~\ref{ass:sur}. Therefore $0 \in Q(\phi|\phi^\ast)$ if and only if $0 \in F(\phi)$. Deduce from this that $\phi^\ast$ is a global minimizer of $Q(\phi|\phi^\ast)$ if and only if $\phi^\ast$ is a global minimizer of $F(\phi)$. 
Thus (\ref{eqn:opt4.1}) and (\ref{eqn:opt4.2}) can be obtained from (\ref{eqn:clarke3.1}) by direct computation.
\qed

\noindent{\textbf{Proof of Lemma~\ref{lem:lem4}}}
                                       
The Dini derivative of $p_\lambda(|\theta|)$ for direction $\varepsilon$ at $\theta^{(0)}$ can be computed to be: 
\begin{equation}                                       
    \begin{aligned}
        p'_{\lambda,D}(|\theta^{(0)}|; \varepsilon)
                                       &=\liminf\limits_{\tau\to 0+}\frac{p_\lambda(|\theta^{(0)}+\tau \varepsilon|)-p_\lambda(|\theta^{(0)}|)}{\tau}\\                                       
        &=
    \begin{cases}
        p_\lambda'(|\theta^{(0)}|)\text{sgn}(\theta^{(0)})\varepsilon, &\mbox{if } |\theta^{(0)}| > 0,\\
        p_\lambda'(0+)|\varepsilon|, &\mbox{if } \theta^{(0)} = 0.\\
	\end{cases}
    \end{aligned}
\end{equation}
Suppose the penalty doesn't change as in (\ref{eqn:nomm}). Let $G_D'(\theta^{(0)}|\theta^{(0)}; \varepsilon)$ denote the Dini derivative of $G(\theta|\theta^{(0)})$ for direction $\varepsilon$ at $\theta^{(0)}$. It is trivially true that \begin{equation}\label{eqn:lem4.3}
G_D'(\theta^{(0)}|\theta^{(0)}; \varepsilon)=p'_{\lambda,D}(|\theta^{(0)}|; \varepsilon).
\end{equation} 
If the penalty is majorized as in (\ref{eqn:lla}), we have
\begin{equation*}
	\begin{aligned}
		&G_D'(\theta|\theta^{(0)}; \varepsilon)\\
		=&\liminf\limits_{\tau\to 0+}\frac{G(\theta+\tau \varepsilon|\theta^{(0)})-G(\theta|\theta^{(0)})}{\tau}\\
		=&\liminf\limits_{\tau\to 0+}\frac{\left(|\theta+\tau \varepsilon|-|\theta|\right)p_\lambda'(|\theta^{(0)}|+)}{\tau}\\
		=&
		\begin{cases}
		p_\lambda'(|\theta^{(0)}|+)\text{sgn}(\theta)\varepsilon, &\mbox{if } |\theta| > 0,\\
		p_\lambda'(|\theta^{(0)}|+)|\varepsilon|, &\mbox{if } \theta = 0.
	        \end{cases}
	\end{aligned}
\end{equation*}
Hence we have
\begin{equation*}
	\begin{aligned}
        G_D'(\theta^{(0)}|\theta^{(0)}; \varepsilon)=
		\begin{cases}
            p_\lambda'(|\theta^{(0)}|)\text{sgn}(\theta^{(0)})\varepsilon, &\mbox{if } |\theta^{(0)}| > 0,\\
            p_\lambda'(|\theta^{(0)}|+)|\varepsilon|, &\mbox{if } \theta^{(0)} = 0.
	        \end{cases}
	\end{aligned}
\end{equation*}
        Again (\ref{eqn:lem4.3}) is obtained.
\qed

	\noindent{\textbf{Proof of Theorem~\ref{thm:mmpg}}}

	Consider the second order Taylor expansion
\begin{equation*}
	L(\phi)= L(\phi^{(k)})+(\phi-\phi^{(k)})^\intercal \nabla L(\phi^{(k)})+\frac{1}{2}(\phi-\phi^{(k)})^\intercal \nabla^2L(\tilde \phi)(\phi-\phi^{(k)}).
\end{equation*}
By the construction of (\ref{eqn:thm1.38}), 
	$\ell(\phi|\phi^{(k)})$ majorizes $L(\phi)$
	and 
	$\ell(\phi|\phi^{(k)})$ strictly majorizes $L(\phi)$ if $\omega > \varrho.$
	Thus 
			to prove that $Q(\phi|\phi^{(k)})$ majorizes  or strictly majorizes $F(\phi)$, 
	we only need to show that $G(\beta_j|\beta_j^{(k)})$ majorizes $p_\lambda(|\beta_j|), 1\leq j\leq p$. 
	We consider three cases depending on the construction of $G(\beta_j|\beta_j^{(k)})$:
	\begin{enumerate}[a)]
		\item If $G(\beta_j|\beta_j^{(k)})$ is defined by (\ref{eqn:nomm}), then trivially  $G(\beta_j|\beta_j^{(k)})$ majorizes $p_\lambda(|\beta_j|)$.
		\item If $G(\beta_j|\beta_j^{(k)})$ is defined by (\ref{eqn:lla}),
			with the concavity of $p_\lambda(\beta_j)$ on $(0, \infty)$, it can be easily shown that 
			$G(\beta_j|\beta_j^{(k)})$ majorizes $p_\lambda(|\beta_j|)$.
			See Theorem 1 in \citet{hunter2005variable} or Theorem 2.1 in \citet{schifano2010majorization}.
		\item If $G(\beta_j|\beta_j^{(k)})$ is defined by (\ref{eqn:lqa}),  
			by Proposition 3.1 in \citet{hunter2005variable},
			$G(\beta_j|\beta_j^{(k)})$ majorizes $p_\lambda(|\beta_j|)$.
	\end{enumerate}
	In summary, we conclude that $Q(\phi|\phi^{(k)})$ majorizes $F(\phi)$, and the inequality in (\ref{eqn:mmp}) holds if $\omega > \varrho$.
	\qed

\noindent{\textbf{Proof of Theorem~\ref{thm:mm2}}}

From Theorem~\ref{thm:mmpg} we have 
\begin{equation*}
	F(\phi^{(k)})=Q(\phi^{(k)}|\phi^{(k)}) \geq Q(\phi^{(k+1)}|\phi^{(k)}) \geq F(\phi^{(k+1)}).
\end{equation*}
We restate (\ref{eqn:thm1.38}) below for convenience:
\begin{equation*}
	\ell(\phi|\phi^{(k)})= L(\phi^{(k)}) + (\phi-\phi^{(k)})^\intercal\nabla L(\phi^{(k)}) + \frac{\omega}{2}(\phi-\phi^{(k)})^\intercal W(\phi-\phi^{(k)}).
\end{equation*}
We prove through an adaption of similar arguments in \citet{lange2016mm} that 
Algorithm~\ref{alg:alg1} with the quadratic upper bound surrogate function 
satisfies the following property:
\begin{equation}\label{eqn:summa}
	Q(\phi|\phi^{(k)}) - Q(\phi^{(k+1)}|\phi^{(k)}) \geq Q(\phi|\phi^{(k+1)}) - F(\phi).
\end{equation}

Denote
\begin{equation*}
	r(\phi)\triangleq\frac{\omega}{2}\phi^\intercal W\phi-L(\phi),
\end{equation*}
\begin{equation*}\label{eqn:bregman1}
	D_r(\phi|\phi^{(k)})\triangleq r(\phi)-r(\phi^{(k)})-(\phi-\phi^{(k)})^\intercal\nabla r(\phi^{(k)}).
\end{equation*}
Then we have
\begin{equation}\label{eqn:pma}
	\ell(\phi|\phi^{(k)})=L(\phi)+D_r(\phi|\phi^{(k)}),
\end{equation}
\begin{equation*}\label{eqn:bregman2}
	\nabla^2 r(\phi)=\omega W - \nabla^2 L(\phi).
\end{equation*}
With (\ref{eqn:wmat1}) we deduce that $\nabla^2 r(\phi)$ is positive semidefinite. Thus $r(\cdot)$ is convex, see for instance, Theorem 2.1.4 in \citet{nesterov2004introductory}. 
From (\ref{eqn:plik}), (\ref{eqn:plik3}), (\ref{eqn:pma}) and $G(\beta|\beta^{(k)})=p_\lambda(|\beta|)$ for some convex function $p_\lambda(|\beta|)$, we rewrite
\begin{equation}\label{eqn:pma2}
	Q(\phi|\phi^{(k)})=F(\phi)+D_r(\phi|\phi^{(k)}).
\end{equation}
By convexity we have
\begin{equation*}
	D_r(\phi|\phi^{(k)}) \geq 0, \quad D_r(\phi^{(k)}|\phi^{(k)})=0.
\end{equation*}
Since $Q(\phi), F(\phi), D_r(\phi|\phi^{(k)})$ and $r(\phi)$ are all convex with respect to $\phi$, we have
\begin{equation*}
    \partial Q(\phi|\phi^{(k)})=\partial F(\phi) + \{\nabla r(\phi)-\nabla r(\phi^{(k)})\},
\end{equation*}
where $\partial Q(\phi|\phi^{(k)})$ and $\partial F(\phi)$ are subdifferentials. Since $\phi^{(k+1)}$ minimizes $Q(\phi|\phi^{(k)})$, then $0 \in \partial Q(\phi^{(k+1)}|\phi^{(k)})$. Equivalently, for some $q^{(k+1)}\in \partial F(\phi)$ we have
\begin{equation}\label{eqn:eqn30}
	q^{(k+1)} + \nabla r(\phi^{(k+1)})-\nabla r(\phi^{(k)})=0.
\end{equation}
From the definition of subgradient we have
\begin{equation}\label{eqn:convex1}
	F(\phi)-F(\phi^{(k+1)})-(\phi-\phi^{(k+1)})^\intercal q^{(k+1)} \geq 0.
\end{equation}
Combining (\ref{eqn:eqn30}) and (\ref{eqn:convex1}), 
we have
\begin{equation*}\label{eqn:eqn34}
	\begin{aligned}
		& Q(\phi|\phi^{(k)}) - Q(\phi^{(k+1)}|\phi^{(k)})\\
	       =&F(\phi)+D_r(\phi|\phi^{(k)}) - F(\phi^{(k+1)}) - D_r(\phi^{(k+1)}|\phi^{(k)})\\
	       =&F(\phi)+r(\phi)-r(\phi^{(k)})-(\phi-\phi^{(k)})^\intercal \nabla r(\phi^{(k)})\\
	        &-F(\phi^{(k+1)})-r(\phi^{(k+1)})+r(\phi^{(k)})+(\phi^{(k+1)}-\phi^{(k)})^\intercal \nabla r(\phi^{(k)})\\
	       =&F(\phi)-F(\phi^{(k+1)})+r(\phi)-r(\phi^{(k+1)})-(\phi-\phi^{(k+1)})^\intercal \nabla r(\phi^{(k)})\\
	       =&F(\phi)-F(\phi^{(k+1)})+r(\phi)-r(\phi^{(k+1)})-(\phi-\phi^{(k+1)})^\intercal \nabla r(\phi^{(k+1)})\\
		&+(\phi-\phi^{(k+1)})^\intercal\left(\nabla r(\phi^{(k+1)})-\nabla r(\phi^{(k)})\right)\\
	       =&F(\phi)-F(\phi^{(k+1)})-(\phi-\phi^{(k+1)})^\intercal q^{(k+1)}+D_r(\phi|\phi^{(k+1)})\\
	    \geq& D_r(\phi|\phi^{(k+1)})
	\end{aligned}    
\end{equation*}
The last inequality and (\ref{eqn:pma2}) means that (\ref{eqn:summa}) holds. Hence Algorithm~\ref{alg:alg1} is a sequential unconstrained minimization algorithm (SUMMA), and $F(\phi^{(k)})$ converges to $\inf F(\phi)$ as shown in \citet{byrne2018auxiliary}.

We prove the second assertion utilizing Proposition 7.4.1 in \citet{lange2016mm} that states if the SUMMA condition holds, $F(\phi)$ and $Q(\phi|\phi^{(k)})$ are continuous, $Q(\phi|\phi^{(k)})$ is uniformly strongly convex, and $F(\phi)$ achieves its minimum, then every sequence $\phi^{(k)}$ converges to a minimum point of $F(\phi)$.  

Since the objective function $F(\phi)$ is convex, thus continuous. The surrogate $Q(\phi|\phi^{(k)})$ is also continuous. From (\ref{eqn:thm1.38}) we have
\begin{equation*}
	\nabla^2\ell(\phi|\phi^{(k)}) = \omega W \succeq \omega \tau I,
\end{equation*}
where $\tau$ is the smallest eigenvalue of $W$.
Hence $\ell(\phi|\phi^{(k)})$ is uniformly strongly convex with parameter $\omega \tau $ which is independent of $\phi^{(k)}$. Note that $\Lambda(\beta)$ is convex provided $G(\beta_j|\beta_j^{(k)})=p_\lambda(|\beta_j|)$ for some convex penalty $p_\lambda(|\cdot|)$. Denote
    \begin{equation*}
		\begin{array}{cc}
				A=(0_{p\times 1} & I_{p\times p}).
	\end{array}
\end{equation*}
Then $A\phi=\beta$. Denote $\tilde{\Lambda}(\phi)\triangleq\Lambda(A\phi)$.
Since the composition of an affine function $A$ preserves convexity of $\Lambda(\cdot)$, $\tilde{\Lambda}(\phi)$ is convex.
	Therefore $Q(\phi|\phi^{(k)})=
	\ell(\phi|\phi^{(k)})+\tilde\Lambda(\phi) 
	=\ell(\phi|\phi^{(k)})+\Lambda(\beta)$ 
	is uniformly strongly convex in $\phi$.

    We need to show that $F(\phi)$ is coercive. First, consider condition (i). Since $L(\phi)$ is coercive meaning $\norm{\phi} \to \infty$ implies $L(\phi) \to \infty$, then $F(\phi)$ is coercive. Next, consider condition (ii). Since $\beta_0$ is bounded, for instance, $\beta_0=0$, i.e., no intercept, we deduce from $\norm\phi \to \infty$ that $\norm\beta \to \infty$, which in turn implies $\Lambda(\beta) \to \infty$ by Assumption~\ref{ass:pen}. Additionally, since $L(\phi)$ is bounded below, then $F(\phi)$ is coercive.  In conclusion, with either (i) or (ii), $F(\phi)$ is coercive. Combining with continuity, hence $F(\phi)$ achieves its minimum \citep[Proposition A.8]{bertsekas1999nonlinear}. 

	In summary, all conditions hold for Proposition 7.4.1 in \citet{lange2016mm}, thus every sequence $\phi^{(k)}$ converges to a minimum point of $F(\phi)$. 
	%
	\qed
\begin{lemma}\label{lem:lem4}
    Suppose Assumption~\ref{ass:pen} holds for $p_\lambda(|\theta|)$, and $G(\theta|\theta^{(0)})$ is given by (\ref{eqn:nomm}) or (\ref{eqn:lla}). Then the Dini derivative of $G(\theta|\theta^{(0)})$ for direction $\varepsilon$ at $\theta^{(0)}$ is the same as that of $p_\lambda(|\theta|)$ given by:
\begin{equation*}\label{eqn:lem4.1}
    G_D'(\theta|\theta^{(0)}; \varepsilon)|_{\theta=\theta^{(0)}} =
	\begin{cases}
        p_\lambda'(|\theta^{(0)}|)\text{sgn}(\theta^{(0)})\varepsilon, &\mbox{if } |\theta^{(0)}| > 0,\\
        p_\lambda'(0+)|\varepsilon|, &\mbox{if } \theta^{(0)} = 0.\\
	\end{cases}
\end{equation*}
\end{lemma}

\noindent{\textbf{Proof of Theorem~\ref{thm:mmpg1}}}
			
To show convergence, we utilize the following assertions:
\begin{enumerate}[(i)]
	\item The surrogate function $Q(\phi|\phi^{(k)})$ is continuous in $(\phi, \phi^{(k)})$. 
    \item The Dini derivative for every direction $\varepsilon$ at $\phi^{(k)}$ is equivalent between $Q(\phi|\phi^{(k)})$ and $F(\phi)$:
	\begin{equation}\label{eqn:mmpde}
		Q'_D(\phi|\phi^{(k)};\varepsilon)|_{\phi=\phi^{(k)}}=F'_D(\phi^{(k)};\varepsilon)\ \forall\ \varepsilon.
\end{equation}
\end{enumerate}
The first term $\ell(\phi|\phi^{(k)})$ is jointly continuous in $(\phi, \phi^{(k)})$ by assumption, and the remaining terms in (\ref{eqn:plik3}) are also jointly continuous. Hence assertion (i) holds.
We prove the second assertion.
Denote $\varepsilon=(\varepsilon_0, ..., \varepsilon_p)^\intercal$, we then compute
\begin{equation*}
	\begin{aligned}
        F'_D(\phi^{(k)}; \varepsilon)&=\liminf\limits_{\tau\to 0+}\frac{F(\phi^{(k)}+\tau \varepsilon)-F(\phi^{(k)})}{\tau}\\
                                       &=\varepsilon^\intercal\nabla L(\phi^{(k)}) + 
        \sum_{j=1}^p\left\{\lambda(1-\alpha)\beta_j^{(k)} \varepsilon_j + \alpha p_{\lambda, D}'(\beta_j^{(k)}; \varepsilon_j)\right\},\\
	\end{aligned}
\end{equation*}
where $p_{\lambda, D}'(\beta_j^{(k)}; \varepsilon_j)$ is the Dini directional derivative of $p_\lambda(|\beta_j|)$ for direction $\varepsilon_j$ at $\beta_j^{(k)}$.
On the other hand, we have
\begin{equation*}
	\begin{aligned}
        Q'_D(\phi|\phi^{(k)}; \varepsilon)_{\phi=\phi^{(k)}}&=\liminf\limits_{\tau\to 0+}\frac{Q(\phi^{(k)}+\tau \varepsilon)-Q(\phi^{(k)})}{\tau}\\
                                       &=\varepsilon^\intercal\nabla\ell(\phi^{(k)}|\phi^{(k)})+
                                         \sum_{j=1}^p\left\{\lambda(1-\alpha)\beta_j^{(k)} \varepsilon_j + \alpha G_D'(\beta_j^{(k)}|\beta_j^{(k)}; \varepsilon_j)
\right\}.
	\end{aligned}
\end{equation*}
By Lemma~\ref{lem:lem4} we have
\begin{equation}\label{eqn:thm7.4}
    p_{\lambda, D}'(\beta_j^{(k)}; \varepsilon_j)=G_D'(\beta_j^{(k)}|\beta_j^{(k)}; \varepsilon_j).
\end{equation}
By Assumption~\ref{ass:sur} we have
\begin{equation}\label{eqn:num47}
	\nabla\ell(\phi^{(k)}|\phi^{(k)})=\nabla L(\phi^{(k)}).
\end{equation}
We deduce from (\ref{eqn:thm7.4}) and (\ref{eqn:num47}) that (\ref{eqn:mmpde}) holds.

Theorem~\ref{thm:mmpg} and assertions (i)-(ii) together imply that Assumption 1 in \citet{razaviyayn2013unified} is satisfied. Thus invoking Theorem 1 in \citet{razaviyayn2013unified} concludes that
			every limit point of the iterates generated by Algorithm~\ref{alg:alg1} is a Dini stationary point of the problem minimizing $F(\phi)$.
\qed

	\noindent{\textbf{Proof of Theorem~\ref{thm:mm3}}}

Consider the following regularity conditions:
\begin{enumerate}[R1.]
\item The objective function $F(\phi)$ is locally Lipschitz continuous and coercive. The set of Clarke stationary points $S$ of $F(\phi)$ is a finite set.
\item $Q(\phi|\phi^{(k)})$ strictly majorizes $F(\phi)$ at $\phi^{(k)}$ for all $\phi, \phi^{(k)}$. 
\item $Q(\phi|\phi^{(k)})$ is continuous for $(\phi, \phi^{(k)})$ and locally Lipschitz for $\phi$ near $\phi^{(k)}$.
\item Denote $M(\phi^{(k)})\triangleq\argmin_\phi Q(\phi|\phi^{(k)})$, then $M(\phi^{(k)})$ is a singleton set consisting of one bounded vector for each $\phi^{(k)}$.
\end{enumerate}
The above conditions lead to the desired result \citep{schifano2010majorization, chi2014robust}. In particular, each limit point $\tilde\phi$ of $\phi^{(k)}$ is a fixed point such that $M(\tilde\phi)=\tilde\phi$ \citep[Theorem A.3]{schifano2010majorization}, and the set of fixed points coincides with the Clarke stationary points of $F(\phi)$ \citep[Proposition A.8]{schifano2010majorization}. We now check each condition.
By assumption $L(\phi)$ is locally Lipschitz continuous, Assumption~\ref{ass:pen} implies that $p_\lambda(|\beta_j)$ is locally Lipschitz continuous due to a bounded derivative at every point including the origin, and $\beta_j^2$ is continuously differentiable thus locally Lipschitz continuous. As a result, $F(\phi)$ is locally Lipschitz continuous.
With the same arguments in the proof of Theorem~\ref{thm:mm2}, $F(\phi)$ is coercive since 
$L(\phi)$ is coercive or bounded below with $\beta_0$ bounded.
With the additional assumption that the set of Clarke stationary points $S$ of $F(\phi)$ is a finite set, R1 is satisfied. 
Since $\ell(\phi|\phi^{(k)})$ strictly majorizes $L(\phi)$ at $\phi^{(k)}$,
and $G(\beta_j|\beta_j^{(k)})$ majorizes $p_\lambda(|\beta_j|)$ at $\beta_j^{(k)}$, 
then $Q(\phi|\phi^{(k)})$ strictly majorizes $F(\phi)$ at $\phi^{(k)}$.
R2 is met.
As a sum of continuous functions on $(\phi, \phi^{(k)})$, the penalized majorization $Q(\phi|\phi^{(k)})$ is continuous. 
The surrogate penalty $\tilde\Lambda(\phi)$ is convex and finite, thus locally Lipschitz continuous. Since $\ell(\phi|\phi^{(k)})$ is also locally Lipschitz, $Q(\phi|\phi^{(k)})$ defined in (\ref{eqn:plik3}) is locally Lipschitz continuous. Thus R3 is met. 
By assumption $\ell(\phi|\phi^{(k)})$ is strongly convex. Then $Q(\phi|\phi^{(k)})$ is strongly convex because it is
a sum of strongly convex function $\ell(\phi|\phi^{(k)})$ and convex function $\tilde\Lambda(\phi)$. Hence $Q(\phi|\phi^{(k)})$ has at most one global minimizer. 
Since $Q(\phi|\phi^{(k)}) \geq F(\phi)$ and $F(\phi)$ is coercive as shown above, then $Q(\phi|\phi^{(k)})$ is coercive. 
Combining with continuity, $Q(\phi|\phi^{(k)})$ achieves its minimum. Together, 
$Q(\phi|\phi^{(k)})$ has one and only one global minimizer. R4 is met. 
\qed
\subsection{Additional results in simulation and data analysis}
\begin{table}[!htbp]
  \caption{
      Mean number of selected variables and standard deviations (in parentheses) for v percentage of outliers in Example 1. 
  \label{ex7nvar}} 
  \begin{center}
    \begin{tabular}{lllll}
      \hline\hline
      \multicolumn{1}{l}{Method}&\multicolumn{1}{c}{v=0}&\multicolumn{1}{c}{v=5}&\multicolumn{1}{c}{v=10}&\multicolumn{1}{c}{v=20}\tabularnewline
      \hline
      LS          &8       &8       &8       &8    \tabularnewline
      LSLASSO    &5.9(1.5)&4.2(2.3)&4.9(2.1)&5.5(1.7)\tabularnewline
      LSSCAD    &4.0(1.4) &3.6(2.0)&4.1(1.9)&5.5(1.7)\tabularnewline
      LSMCP    &3.9(1.3)  &3.2(2.1)&3.6(2.0)&5.5(1.7)\tabularnewline
      HuberLASSO &5.2(1.1)&5.3(1.2)&5.3(1.2)&5.2(1.2)\tabularnewline
      ClossRLASSO&5.9(1.4)&5.9(1.5)&6.0(1.5)&5.9(1.4)\tabularnewline
        ClossRSCAD &4.1(1.4)&4.3(1.7)&4.1(1.5)&\textbf{4.1}(1.5)\tabularnewline
        ClossRMCP  &\textbf{3.9}(1.3)&\textbf{4.0}(1.6)&\textbf{3.9}(1.4)&3.9(1.4)\tabularnewline
      \hline
  \end{tabular}\end{center}

\end{table}

\begin{table}[!htbp]
  \caption{
      Mean sensitivity and standard deviations (in parentheses) for v percentage of outliers in Example 1. 
  \label{ex7sen}} 
  \begin{center}
    \begin{tabular}{lllll}
      \hline\hline
      \multicolumn{1}{l}{Method}&\multicolumn{1}{c}{v=0}&\multicolumn{1}{c}{v=5}&\multicolumn{1}{c}{v=10}&\multicolumn{1}{c}{v=20}\tabularnewline
      \hline
      LS          &1       &1       &1       &1    \tabularnewline
      LSLASSO    &1(0)&0.70(0.28)&0.84(0.21)&0.95(0.12)\tabularnewline
        LSSCAD    &1(0) &0.63(0.28)&0.77(0.24)&0.87(0.19)\tabularnewline
        LSMCP    &1(0)  &0.57(0.28)&0.71(0.27)&0.84(0.20)\tabularnewline
        HuberLASSO &1(0)&1(0)&1(0)&1(0)\tabularnewline
        ClossRLASSO&1(0)&1(0)&1(0)&1(0)\tabularnewline
        ClossRSCAD &1(0)&1(0)&1(0)&1(0)\tabularnewline
        ClossRMCP  &1(0)&1(0)&1(0)&1(0)\tabularnewline
      \hline
  \end{tabular}\end{center}

\end{table}

\begin{table}[!htbp]
  \caption{
      Mean specificity and standard deviations (in parentheses) for v percentage of outliers in Example 1. 
  \label{ex7spc}} 
  \begin{center}
    \begin{tabular}{lllll}
      \hline\hline
      \multicolumn{1}{l}{Method}&\multicolumn{1}{c}{v=0}&\multicolumn{1}{c}{v=5}&\multicolumn{1}{c}{v=10}&\multicolumn{1}{c}{v=20}\tabularnewline
      \hline
        LS          &0       &0       &0       &0    \tabularnewline
      LSLASSO    &0.42(0.29)  &0.57(0.34)&0.53(0.34)&0.47(0.32)\tabularnewline
        LSSCAD   &0.80(0.27)  &0.66(0.31)&0.64(0.30)&0.64(0.32)\tabularnewline
        LSMCP    &0.82(0.26)  &0.70(0.31)&0.70(0.29)&0.69(0.30)\tabularnewline
        HuberLASSO &0.55(0.22)&0.54(0.24)&0.54(0.25)&0.55(0.24)\tabularnewline
        ClossRLASSO&0.41(0.29)&0.41(0.30)&0.40(0.29)&0.41(0.28)\tabularnewline
        ClossRSCAD &0.78(0.28)&0.75(0.33)&0.78(0.31)&0.78(0.30)\tabularnewline
        ClossRMCP  &0.82(0.26)&0.80(0.32)&0.82(0.28)&0.81(0.28)\tabularnewline
      \hline
  \end{tabular}\end{center}

\end{table}

\begin{table}[!b]
    \caption{Mean number of selected variables and standard deviations (in parentheses) for v pencentage of outliers in Example 2.}
    \label{tab:ex1nvar}
	\begin{center}
		\begin{tabular}{lllll}
			\hline\hline
			\multicolumn{1}{l}{Method}&\multicolumn{1}{c}{v=0}&\multicolumn{1}{c}{v=5}&\multicolumn{1}{c}{v=10}&\multicolumn{1}{c}{v=20}\tabularnewline
\hline
            Bayes &2&2&2&2\tabularnewline
            Logis&20&20&20&20\tabularnewline
            LogisLASSO&2.4(0.7)&3.0(1.7)&3.2(1.9)&3.7(2.1)\tabularnewline
            LogisSCAD&2.0(0.0)&2.3(1.1)&2.9(1.8)&4.1(2.1)\tabularnewline
            LogisMCP&2.0(0.1)&2.5(1.1)&2.9(1.8)&3.3(1.7)\tabularnewline
            ClossLASSO&2.6(0.7)&2.6(0.7)&5.0(2.3)&4.8(2.2)\tabularnewline
            ClossSCAD&2.0(0)&2.0(0)&2.0(0)&2.0(0.1)\tabularnewline
            ClossMCP&2.0(0)&2.0(0)&2.0(0.2)&2.1(0.3)\tabularnewline
            GlossLASSO&4.1(1.6)&4.2(1.7)&4.3(1.8)&4.9(2.5)\tabularnewline
            GlossSCAD&2.0(0.1)&\textbf{2.0}(0.1)&\textbf{2.0}(0.1)&2.3(0.7)\tabularnewline
            GlossMCP&2.0(0.2)&2.1(0.2)&2.1(0.4)&2.2(0.7)\tabularnewline
            QlossLASSO&5.1(2.7)&5.2(2.7)&5.2(2.9)&5.3(3.0)\tabularnewline
            QlossSCAD&\textbf{2.0}(0)&2.0(0)&2.0(0)&\textbf{2.0}(0.1)\tabularnewline
            QlossMCP&2.0(0.1)&2.0(0.2)&2.0(0)&2.1(0.2)\tabularnewline
\hline
\hline
\end{tabular}\end{center}

\end{table}

\begin{table}[!tbp]
    \caption{Mean test errors and standard deviations (in parentheses) for v pencentage of outliers in Example 3.}
    \label{tab:ex2err}
	\begin{center}
		\begin{tabular}{lllll}
			\hline\hline
			\multicolumn{1}{l}{Method}&\multicolumn{1}{c}{v=0}&\multicolumn{1}{c}{v=5}&\multicolumn{1}{c}{v=10}&\multicolumn{1}{c}{v=20}\tabularnewline
\hline
            Bayes &0&0.05&0.1&0.2\tabularnewline
            Logis&0.0224(0.0052) &0.1166(0.0069)&0.1780(0.0068) &0.2831(0.0070)\tabularnewline
            LogisLASSO&0.0084(0.0083) &0.0626(0.0101)&0.1153(0.0106)&0.2190(0.0136)\tabularnewline
            LogisSCAD&0.0037(0.0042)&0.0591(0.0079)&0.1123(0.0090)&0.2281(0.0133)\tabularnewline
            LogisMCP&0.0036(0.0042)&0.0591(0.0079)&0.1125(0.0092)&0.2176(0.0128)\tabularnewline
            ClossLASSO&0.0041(0.0042)&0.0539(0.0039)&0.1041(0.0041)&0.2054(0.0044)\tabularnewline
            ClossSCAD&0.0048(0.0035)&0.0544(0.0034)&0.1041(0.0034)&0.2038(0.0030)\tabularnewline
            ClossMCP &0.0048(0.0035)&0.0544(0.0034)&0.1041(0.0034)&0.2043(0.0057)\tabularnewline
            GlossLASSO&0.0050(0.0047)&0.0554(0.0047)&0.1051(0.0047)&0.2071(0.0059)\tabularnewline
            GlossSCAD&0.0016(0.0018)&0.0522(0.0029)&0.1018(0.0018)&0.2043(0.0052)\tabularnewline
            GlossMCP&\textbf{0.0016}(0.0015)&0.0530(0.0033)&0.1040(0.0036)&0.2052(0.0041)\tabularnewline
            QlossLASSO&0.0049(0.0045)&0.0555(0.0048)&0.1050(0.0048)&0.2067(0.0057)\tabularnewline
            QlossSCAD&0.0016(0.0016)&\textbf{0.0520}(0.0018)&\textbf{0.1018}(0.0016)&\textbf{0.2019}(0.0018)\tabularnewline
            QlossMCP&0.0017(0.0017)&0.0523(0.0027)&0.1027(0.0031)&0.2033(0.0042)\tabularnewline
\hline
\hline
\end{tabular}\end{center}

\end{table}

\begin{table}[!tbp]
    \caption{Mean number of selected variables and standard deviations (in parentheses) for v pencentage of outliers in Example 3.}
    \label{tab:ex2nvar}
	\begin{center}
		\begin{tabular}{lllll}
			\hline\hline
			\multicolumn{1}{l}{Method}&\multicolumn{1}{c}{v=0}&\multicolumn{1}{c}{v=5}&\multicolumn{1}{c}{v=10}&\multicolumn{1}{c}{v=20}\tabularnewline
\hline
            Bayes &2&2&2&2\tabularnewline
            Logis&40&40&40&40\tabularnewline
            LogisLASSO&2.1(0.4)&3.3(1.6)&3.4(1.8)&4.1(3.3)\tabularnewline
            LogisSCAD&2.0(0)&2.9(1.4)&3.2(2.0)&3.8(2.0)\tabularnewline
            LogisMCP&2.0(0)&2.8(1.2)&3.0(1.4)&3.4(1.8)\tabularnewline
            ClossLASSO&2.2(0.4)&2.2(0.4)&2.4(0.6)&2.6(0.8)\tabularnewline
            ClossSCAD&2.0(0)&2.0(0)&2.0(0)&2.0(0)\tabularnewline
            ClossMCP&2.0(0)&2.0(0)&2.0(0)&2.0(0.1)\tabularnewline
            GlossLASSO&2.7(1.3)&2.7(0.8)&2.7(1.4)&2.6(0.9)\tabularnewline
            GlossSCAD&2.1(0.3)&2.1(0.3)&2.0(0.1)&2.4(0.7)\tabularnewline
            GlossMCP&\textbf{2.0}(0.1)&2.2(0.5)&2.4(0.7)&2.7(0.8)\tabularnewline
            QlossLASSO&2.4(0.7)&2.5(0.7)&2.5(0.7)&2.6(0.8)\tabularnewline
            QlossSCAD&2.0(0.2)&\textbf{2.0}(0.2)&\textbf{2.0}(0)&\textbf{2.0}(0)\tabularnewline
            QlossMCP&2.1(0.2)&2.1(0.4)&2.1(0.3)&2.2(0.4)\tabularnewline
\hline
\hline
\end{tabular}\end{center}

\end{table}

\begin{table}[!tbp]
\caption{Variable selection for $x_1$ or $x_2$ for v percentage of outliers in Example 3.\label{tab:ex2_var}} 
\begin{center}
\begin{tabular}{lrrrr}
\hline\hline
\multicolumn{1}{l}{Method}&\multicolumn{1}{c}{v=0}&\multicolumn{1}{c}{v=5}&\multicolumn{1}{c}{v=10}&\multicolumn{1}{c}{v=20}\tabularnewline
\hline
LogisLASSO&$0$&$0$&$1$&$3$\tabularnewline
LogisSCAD&$0$&$0$&$1$&$5$\tabularnewline
LogisMCP&$0$&$0$&$1$&$4$\tabularnewline
ClossLASSO&$0$&$0$&$2$&$3$\tabularnewline
ClossSCAD&$0$&$0$&$0$&$0$\tabularnewline
ClossMCP&$0$&$0$&$0$&$0$\tabularnewline
GlossLASSO&$0$&$0$&$2$&$0$\tabularnewline
GlossSCAD&$2$&$1$&$0$&$1$\tabularnewline
GlossMCP&$0$&$0$&$1$&$3$\tabularnewline
QlossLASSO&$0$&$0$&$1$&$0$\tabularnewline
QlossSCAD&$0$&$1$&$0$&$0$\tabularnewline
QlossMCP&$0$&$1$&$0$&$0$\tabularnewline
\hline
\end{tabular}\end{center}
\end{table}

\begin{table}[!tbp]
    \caption{Mean number of selected variables and standard deviations (in parentheses) for v pencentage of outliers in the breast cancer data.}\label{tab:cancernvar} 
  \begin{center}
    \begin{tabular}{lllll}
      \hline\hline
      \multicolumn{1}{l}{Method}&\multicolumn{1}{c}{v=0}&\multicolumn{1}{c}{v=5}&\multicolumn{1}{c}{v=10}&\multicolumn{1}{c}{v=15}\tabularnewline
      \hline
      ~~LogisLASSO&11.7(3.6) &    16.5(5.2)    &  18.6(7.3)&    19.4(7.7)\tabularnewline
      ~~LogisSCAD&9.1(4.3) &    15.3(5.8)    &  18.5(7.3)&    19.1(7.6)\tabularnewline
      ~~LogisMCP&8.4(3.3) &    11.9(4.5)    &  14.1(6.7)&    15.1(6.7)\tabularnewline
        ~~ClossLASSO&\textbf{5.2}(2.3) &    5.1(2.6)    &  6.0(3.1)&    7.3(4.3)\tabularnewline
        ~~ClossSCAD&1.5(1.0)     &1.5(1.1)      &1.8(1.4)     & \textbf{2.8}(2.4)\tabularnewline
        ~~ClossMCP&1.0(0)&     \textbf{1.0}(0)    &  \textbf{1.0}(0.1)&      1.0(0.9)\tabularnewline
      ~~GlossLASSO&5.2(2.2) &    6.0(2.9)    &  7.6(4.6)&    10.2(5.3)\tabularnewline
      ~~GlossSCAD&1.0(0)     &1.0(0)      &1.0(0)     & 2.1(4.4)\tabularnewline
      ~~GlossMCP&1.1(0.7)&     1.4(1.0)    &  2.1(2.8)&      4.7(4.9)\tabularnewline
      ~~QlossLASSO&5.8(2.3) &    6.2(2.8)    &  7.9(5.0)&    10.7(5.9)\tabularnewline
      ~~QlossSCAD&1.0(0)     &1.0(0)      &1.0(0)     & 1.8(3.6)\tabularnewline
      ~~QlossMCP&1.3(0.8)&     1.5(1.2)    &  2.2(3.0)&      4.9(4.9)\tabularnewline
      \hline
  \end{tabular}\end{center}

\end{table}

\end{appendices}

\end{document}